\documentclass[11pt]{article}
\usepackage{geometry}                
\usepackage{graphicx}
\usepackage{amssymb}
\usepackage{epstopdf}
\usepackage{amsmath, amsthm}
\usepackage{longtable}
\newtheorem{theorem}{Theorem}[section]
\newtheorem{proposition}[theorem]{Proposition}
\newtheorem{lemma}[theorem]{Lemma}
\newtheorem{corollary}[theorem]{Corollary}
\newcommand{\1}{\mathbf{1}}
\newcommand{\Tr}{\mathrm{Tr}}
\newcommand\T{\rule[-1.4ex]{0.0pt}{3.8ex}}

\begin{document}

\begin{center}{\Large \textbf{Pfaffian point process for the Gaussian real generalised eigenvalue problem}}\\ \vspace{36pt}{\large Peter J. Forrester and Anthony Mays}\\ \vspace{18pt}\textit{Department of Mathematics and Statistics, University of Melbourne, Victoria 3010, Australia}
\end{center}

\vspace{36pt}

\begin{abstract}
The generalised eigenvalues for a pair of $N\times N$ matrices $(X_1,X_2)$ are defined as the solutions of the equation $\det (X_1-\lambda X_2)=0$, or equivalently, for $X_2$ invertible, as  the eigenvalues of $X_2^{-1}X_1$. We consider Gaussian real matrices $X_1,X_2$, for which the generalised eigenvalues have the rotational invariance of the half-sphere, or after a fractional linear transformation, the rotational invariance of the unit disk.  In these latter variables we calculate the joint eigenvalue probability density function, the probability $p_{N,k}$ of finding $k$ real eigenvalues, the densities of real and complex eigenvalues (the latter being related to an average over characteristic polynomials), and give an explicit Pfaffian formula for the higher correlation functions $\rho_{(k_1,k_2)}$. A limit theorem for $p_{N,k}$ is proved, and the scaled form of $\rho_{(k_1,k_2)}$ is shown to be identical to the analogous limit for the correlations of the eigenvalues of real Gaussian matrices. We show that these correlations satisfy sum rules characteristic of the underlying two-component Coulomb gas.
\end{abstract}
\newpage

\section{Introduction}

The general topic of our study is the statistical properties of the eigenvalues of $N \times N$ non-symmetric random matrices with real entries. Such matrices will, in general, have both real and complex eigenvalues. In the case that the entries of the matrix are independently chosen as standard Gaussians --- referred to as the real Ginibre ensemble after \cite{Gi65} --- a result of Edelman \textit{et al} \cite{eks1994} tells us that the expected number of real eigenvalues is asymptotically equal to $\sqrt{2N/\pi}$. Numerical evidence presented in the same paper indicates that this asymptotic value persists with the standard Gaussian replaced by any distribution of zero mean and unit standard deviation.

Similar properties hold true of the generalised eigenvalues of a pair of $N\times N$ random matrices $(X_1,X_2)$ with standard Gaussian entries. The generalised eigenvalues $\lambda$ are specified as the solutions of $\det(X_1-\lambda X_2)=0$, or equivalently, as the eigenvalues of $X_2^{-1}X_1$. The statistical properties of the $\lambda$ are the specific concern of this paper. Our starting point will be to first establish the joint matrix distribution of $Y=X_2^{-1}X_1$, which we show in Proposition \ref{prop:elementjpdf}  to be the matrix Cauchy distribution
\begin{eqnarray}
\label{eqn:elementjpdf} \mathcal{P}(Y)&=&\pi^{-N^2/2}\hspace{3pt}
\prod_{j=0}^{N-1}
\frac{\Gamma((N+1)/2+j/2)}{\Gamma((j+1)/2)}\hspace{3pt}\mathrm{det}(\mathbf{1}_N+YY^T)^{-N}.
\end{eqnarray}

Studies into the statistical properties of the $\lambda$ were initiated in \cite{eks1994} using a different logic. It was shown, for example, that the expected number $E_N$ of real (generalised) eigenvalues  has the exact evaluation 
\begin{eqnarray}
\label{eqn:eks_result}
E_N=\frac{\sqrt{\pi}\hspace{2pt}\Gamma((N+1)/2)}{\Gamma(N/2)} \mathop{\sim}\limits_{N \to
\infty} \sqrt{\pi N \over 2}.
\end{eqnarray}
To derive (\ref{eqn:eks_result}), the generalised  eigenvalue problem in the case of $X_1,X_2$ having standard Gaussian elements was placed in the context of integral geometry.

First the pair of matrices $(X_1,X_2)$ can be regarded as two vectors in $\mathbb{R}^{N^2}$ and the corresponding plane spanned by these vectors intersects the sphere $S^{N^2-1}$ to give a great circle. The real generalised eigenvalues correspond to the intersection  of this great circle with the set $\Delta_N$ of all $N \times N$ singular matrices $X$ such that $\mathrm{Tr} XX^T=1$ (thus choose $X=c(X_1-\lambda X_2)$ for suitable $c$). With $X_1,X_2$ having standard Gaussian entries, the great circle has uniform measure, so the expected number of real eigenvalues is equal to the expected number of intersections of $\Delta_N$ with a random great circle.

Another feature of the random generalised eigenvalue problem studied in \cite{eks1994} is the density $\rho_{(1)}(\lambda)$ of real generalised eigenvalues. By writing $\lambda =\mathrm{tan} \hspace{2pt}\theta$ the generalised eigenvalue equation reads $\det (\mathrm{cos}\theta X_1-\mathrm{sin}\theta X_2)=0$. Using the fact that a pair of standard Gaussians $(x_1,x_2)$ is, as a distribution in the plane, invariant under rotation, it was noted that $(\mathrm{cos} \theta,\mathrm{sin} \theta)$ must be distributed uniformly on the unit circle, and so
\begin{eqnarray}
\label{eqn:rho}
\rho_{(1)}(\lambda)=\frac{1}{\pi}\frac{E_N}{1+\lambda^2}.
\end{eqnarray}

The transformation $\lambda =\mathrm{tan} \hspace{2pt}\theta$ is the stereographic projection of the real line on to a great circle of the sphere. From the famous circle theorem \cite{Gi84,Ba97,tvk} relating to eigenvalue densities of large $N \times N$ matrices with general iid entries drawn from any distribution with mean zero and fixed $\sigma$, one might similarly expect uniform asymptotic density on the sphere for $A^{-1}B$ where $A,B$ also have iid entries from any distribution with zero mean and fixed $\sigma$ --- a kind of spherical law\footnote{This has recently been proved by
Bordenave \cite{Bo11}.}. Certainly, in the case of complex Gaussian entries, this is true for every value of $N$. Indeed, with
\begin{eqnarray}\label{2'}
C:=-\bar{\beta}B+\bar{\alpha}A, &D:=\alpha B+\beta A,
\end{eqnarray}
where $\alpha,\beta\in\mathbb{C}$ such that $|\alpha|^2+|\beta|^2=1$, one has that $(A,B)$ has the same distribution as $(C,D)$, implying that the corresponding joint eigenvalue distribution is invariant under rotation of the sphere \cite{krishnapur2008}. In the traditions of random matrix theory --- for example the circular ensembles of Dyson --- the present real Gaussian matrices $A^{-1}B$ may be said to form the real spherical ensemble. In \cite{krishnapur2008} the complex Gaussian matrices $A^{-1}B$ were referred to as the (complex) spherical ensemble.

From (\ref{eqn:rho}) and the fact that with $\lambda = \tan \theta$, $d\lambda/(1 + \lambda^2) = d \theta$, we see that as a distribution on the sphere, the density of real eigenvalues is uniform. One can also stereographically project the complex eigenvalues on to the sphere; it will turn out (eq.~(\ref{df}) below) that in the large $N$ limit the total eigenvalue distribution is uniform. Thus, to leading order, the concentration of real eigenvalues on a great circle of the sphere does not affect the overall eigenvalue distribution. This is analogous to the situation for the real Ginibre ensemble, for which the expected number of real eigenvalues is asymptotically $\sqrt{N/2\pi}$, yet the eigenvalue density is to leading order uniform \cite{edelman1997}. Also, we remark that there is an analogy with the random polynomials
\begin{eqnarray}
\label{eqn:rand_polys}
p_{n}(z)=\sum_{p=0}^n {n \choose p}^{1/2}a_p z^p,&a_p\sim N[0,1].
\end{eqnarray}
When stereographically projected onto the sphere there is of order $\sqrt{N}$ zeros on a great circle corresponding to the real axis \cite{EK95}, but for $N$ large the density is asymptotically uniform on the sphere \cite{Mc09}.

\begin{figure}[ht]
\begin{center}
\includegraphics[scale=0.6]{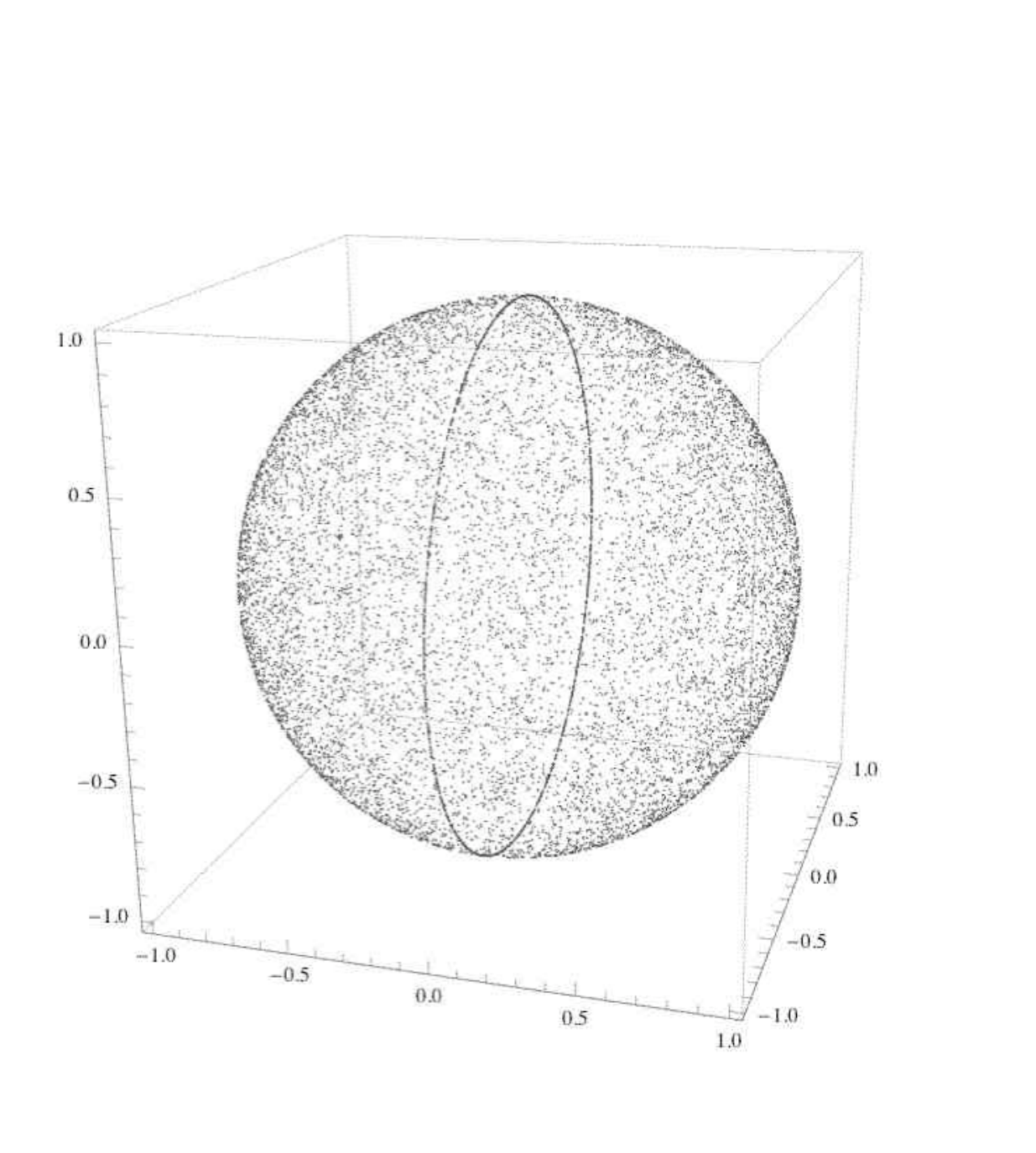}
\end{center}
\caption{A plot of the eigenvalues from 120 realisations of $100\times 100$ matrices with the eigenvalues stereographically projected. The great circle of real eigenvalues can be clearly seen.}
\end{figure}

So in summary, from the pioneering study of the Gaussian real generalised eigenvalue problem of \cite{eks1994}, we know the expected number of real eigenvalues and the density of real
eigenvalues. Beginning with (\ref{eqn:elementjpdf}), this statistical knowledge will be greatly extended. 

An essential ingredient in our analysis is to conformally map from the real line and upper half-plane (\textit{i.e.} the domain of the independent eigenvalues of $Y$) to the boundary of the unit disk by the fractional linear transformation
\begin{eqnarray}
\label{7'} z = {1 \over i} {w -1 \over w+1}.
\end{eqnarray}
For $\lambda$ a real eigenvalue, and with $e:=e^{i\theta}$, (\ref{7'}) reads
\begin{eqnarray}\label{14.2}
\lambda&=&\mathrm{tan}\frac{\theta}{2}=\frac{1}{i}\left(1-\frac{2}{e+1}\right).
\end{eqnarray}
A key feature is that the corresponding real eigenvalue density is uniform in $\theta$
\begin{eqnarray}
\label{eqn:rhor} \rho_{(1)}^r(\theta)=\frac{E_N}{2\pi},
\end{eqnarray}
as follows from (\ref{eqn:rho}). In terms of (\ref{7'}) and (\ref{14.2}) we show that the explicit form of the joint eigenvalue probability density function (jpdf) for $k$ real eigenvalues is
\begin{eqnarray}
\label{eqn:q(y)} \mathcal{Q}(Y)=A_{k,N}\prod_{j=1}^k\tau(e_j)\prod_{s=k+1}^{(N+k)/2}\frac{1}{|w_s|^2}\tau(w_s)\tau\left(\frac{1}{\bar{w}_s}\right)\Delta\left(\mathbf{e},\mathbf{w},\mathbf{\frac{1}{\bar{w}}}\right),
\end{eqnarray}
with $\mathbf{e}=\{e_1,...,e_k\}, \mathbf{w}=\{w_1,\bar{w}_1,...,w_{(N-k)/2},\bar{w}_{(N-k)/2}\}$ and
\begin{eqnarray}
\nonumber A_{k,N}&=& \frac{(-1)^{(N-k)/2((N-k)/2-1)/2+(N-k)k/2-k(k-1)/4}}{2^{(N(N-1)+k)/2}}\prod_{j=1}^N\frac{1}{\Gamma(j/2)^2}\\
&&\nonumber \times\Gamma((N+1)/2)^{N/2}\Gamma(N/2+1)^{N/2},\\
\nonumber \tau(x)&=&\left( \frac{1}{x}\right)^{(N-1)/2}\left[\frac{1}{\sqrt{\pi}}\int_{\frac{|x|^{-1}-|x|}{2}}^{\infty}\frac{dt}{\left(1+t^2\right)^{N/2+1}}\right]^{1/2},
\end{eqnarray}
\begin{eqnarray}
\nonumber \Delta\left(\mathbf{e},\mathbf{w},\mathbf{\frac{1}{\bar{w}}}\right)&=&\prod_{j<p}(e_p-e_j)\prod_{j=1}^k\prod_{s=k+1}^{(N+k)/2}(w_s-e_j)(\frac{1}{\bar{w}_s}-e_j)\\
\nonumber &&\times \prod_{a<b}(w_b-w_a)\left(\frac{1}{\bar{w}_s}-\frac{1}{\bar{w}_s}\right)\prod_{c,d=k+1}^{(N+k)/2}\left( \frac{1}{\bar{w}_d}-w_c\right),
\end{eqnarray}
which is the content of Proposition \ref{thm:ainvb_jpdf}.

We use (\ref{eqn:q(y)}) to deduce exact statistical properties. As an example, let $p_{N,k}$ denote the probability that there are exactly $k$ real eigenvalues, and set
\begin{eqnarray}\label{3'}
\nonumber Z_N(\xi)=\sum_{k=0}^{N}{}^* \xi^k p_{N,k},
\end{eqnarray}
where the asterisk indicates that the sum over $k$ is restricted to values with the same parity as $N$. In Proposition \ref{prop:gen_fn}, with $N$ even, we show
\begin{eqnarray}\label{3a}
\nonumber Z_N(\xi)&=&\frac{(-1)^{(N/2)(N/2-1)/2}}{2^{N(N-1)/2}}\Gamma((N+1)/2)^{N/2}\Gamma(N/2+1)^{N/2}\\
\label{eqn:Z_N} &&\times\prod_{s=1}^{N}\frac{1}{\Gamma(s/2)^2}\prod_{l=0}^{N/2-1}(\xi^2 \alpha_{l} + \beta_{l}),
\end{eqnarray}
where
\begin{eqnarray}\label{3'a}
\nonumber \alpha_{l}&=&\frac{2\pi}{N-1-4l}\frac{\Gamma((N+1)/2)}{\Gamma(N/2+1)},\\
\beta_{l}&=&\frac{2\sqrt{\pi}}{N-1-4l}\left( 2^N\frac{\Gamma(2l+1)\Gamma(N-2l)}{\Gamma(N+1)}-\sqrt{\pi}\frac{\Gamma((N+1)/2)}{\Gamma(N/2+1)}\right).
\end{eqnarray}
The product form (\ref{3a}) for $Z_N(\xi)$ should be contrasted with the simplest known forms in the case of the real Ginibre ensemble \cite{AK07,FN08}, which are determinants of size $N/2$.

It follows from (\ref{eqn:Z_N}) that
\begin{eqnarray}\label{3.1}
E_N=\left. \frac{\partial}{\partial \xi}Z_N(\xi) \right|_{\xi=1}=\sum_{l=0}^{N/2-1}\frac{2\hspace{1pt}\alpha_l}{\alpha_l+\beta_l},
\end{eqnarray}
where we have used the fact that $Z_N(1)=1$. Substituting (\ref{3'a}) then reclaims (\ref{eqn:eks_result}) with $N$ even. In Corollary \ref{prop:odd_EN_var} we give the odd analogue of (\ref{3.1}) and with (\ref{3'a}) we again reclaim (\ref{eqn:eks_result}). The variance in the distribution of the number of real eigenvalues is, by definition, given by $\sigma_N^2 = \sum_{k=0}^N k^2 p_{N,k} - E_N^2$, which, in terms of the generating function, reads
\begin{equation}\label{6.1}
\sigma_N^2 = {\partial^2 \over \partial \xi^2} Z_N(\xi) \Big |_{\xi = 1} + E_N - E_N^2.
\end{equation}
Substituting (\ref{eqn:Z_N}) and (\ref{3.1}), we then obtain the explicit evaluation
\begin{equation}\label{3.1a}
\sigma_N^2=2E_N - 2\sqrt{\pi}\frac{\Gamma(N-1/2)\Gamma((N+1)/2)^2}{\Gamma(N)\Gamma(N/2)^2}.
\end{equation}
This has large $N$ form $\sigma_N^2 \sim (2-\sqrt{2})E_N$, which is the same as for the real Ginibre ensemble \cite{forrester and nagao2007}.

Recalling (\ref{3'}) we can read off the explicit value of $p_{N,k}$. For $N,k$ odd it is a rational number, while for $N,k$ even it is of the form $\sum_{l=k/2}^{N/2} c_{N,2l} \pi^{2l}$ with $c_{N,2l}$ rational (Proposition \ref{p310}). In particular, for $N = 2$,
\begin{equation}\label{2.2}
p_{2,0} = 1 - {\pi \over 4}, \qquad p_{2,2} = {\pi \over 4},
\end{equation}
and for $N = 3$,
\begin{equation}
p_{3,1} = p_{3,3} = {1 \over 2}.
\end{equation}
As decimals (\ref{2.2}) reads $p_{2,0} \approx 0.785$, $p_{2,2} \approx 0.214$. These values, further approximated to $0.79$ and $0.21$ respectively, have been reported in a simulation of the $N=2$ case \cite{Kr89}. The latter study was motivated by the question of determining the typical rank of a $2 \times 2 \times 2$ array (tensor) (see, e.g., \cite{Ma07}).

The meaning of a tensor in this setting relates to structures $\mathcal A = (a_{ijk}) \in
\mathbb R^{p \times p \times 2}$ represented as the column vector
${\rm vec } \, \mathcal A \in \mathbb R^{4 p^2}$. As reviewed in \cite{KB09}, a point of interest is to find matrices
$U = [ \vec{u}_1 \cdots \vec{u}_R] \in \mathbb R^{p \times R}$,
$V = [ \vec{v}_1 \cdots \vec{v}_R] \in \mathbb R^{p \times R}$ and
$W = [ \vec{w}_1 \cdots \vec{w}_R] \in \mathbb R^{2 \times R}$ such that
$$
{\rm vec} \, \mathcal A = \sum_{r=1}^R \vec{w}_r \otimes \vec{v}_r \otimes \vec{u}_r
$$
for $R$ as small as possible. The positive integer $R$ is referred to the rank.
With both $(a_{ij1}) =: X_1 \in \mathbb R^{p \times p}$ and
$(a_{ij2}) =: X_2 \in \mathbb R^{p \times p}$ random matrices, entries chosen from a continuous
distribution, one has that $R = p$ if all the eigenvalues of $X_1^{-1} X_2$ are real,
and $R = p + 1$ otherwise.

The simple structure of (\ref{eqn:Z_N}) also allows for the computation of the large $N$ form of the distribution of the number $k$ of real eigenvalues. In Proposition \ref{pAB} we prove that it is a standard Gaussian in the scaled variable $(k-E_N)/\sigma_N$. Further, our method of derivation of (\ref{eqn:Z_N}), which involves first computing the exact functional form of the eigenvalue jpdf (\ref{eqn:q(y)}), allows for the result (\ref{eqn:rho}) to be generalised. Thus, in Theorem \ref{thm:correlns} we give a $(k_1 + k_2) \times (k_1 + k_2)$ Pfaffian formula for the exact $(k_1,k_2)$-point correlation function between $k_1$ real eigenvalues and $k_2$ complex eigenvalues. The simplest case beyond (\ref{eqn:rho}) is the density of the complex eigenvalues in terms of the coordinates (\ref{7'}). With $r := |w|$ we find that the complex density $\rho_{(1)}^c(w)$ depends only on $r$,
and is given by
\begin{equation}
\label{eqn:rhoc} \rho_{(1)}^c(w) = {N(N-1) \over 2^{N+1} \pi r^2} \Big ( {1 \over r} + r \Big )^{N-2}
\Big ( {1 \over r} - r \Big )
\int_{r^{-1} - r \over 2}^\infty {dt \over (1 + t^2)^{N/2 + 1}}.
\end{equation}
The densities (\ref{eqn:rhoc}) and (\ref{eqn:rhor}) are related by the sum rule for the total number of eigenvalues
\begin{eqnarray}
\nonumber \int_0^{2\pi}d\theta \hspace{3pt}\rho_{(1)}^r(\theta)+2\int_0^1r \; dr\int_0^{2\pi}d\theta \hspace{3pt}\rho_{(1)}^c(w)&=&N,
\end{eqnarray}
where the factor of 2 is required since $\rho_{(1)}^c$ only refers to one of each complex conjugate pair.

Section 4 continues with  the evaluation of the average of two characteristic polynomials in terms of elements of the correlation kernel. We also compute the scaled limit of the $(k_1,k_2)$-point correlation function, and demonstrate that it agrees with the recently obtained \cite{BS08} scaled correlation function for the bulk eigenvalues in the real Ginibre ensemble. An analogy with a Coulomb gas allows for the formulation of sum rules relating to the screening of the effective charge of a fixed number of eigenvalues, and also allows us to isolate a certain combination of one- and two-point correlations for which the complex moments vanish.

\section{Joint probability density functions}
\subsection{Element distribution}

For $A,B$ matrices of size $N\times N$ taken from Ginibre's real ensemble so that
\begin{equation}\label{eqn:YjpdfAB} 
\mathcal{P}(A)\cdot \mathcal{P}(B)
= (2\pi)^{-N^2}e^{-\frac{1}{2}\mathrm{Tr}(AA^T+BB^T)},
\end{equation}
we wish to express the probability density function $\mathcal{P}(Y)$ of the elements of the matrix $Y=A^{-1}B$ in terms of the elements of $Y$. This we read off from the calculation of $\mathcal{P}(Y)(dY)$, where $(dY)$ is the wedge product of the independent elements of the matrix $dY$.
\begin{proposition}
\label{prop:elementjpdf}
Let $A,B$ be $N\times N$ matrices drawn from the real Ginibre ensemble, and let $Y=A^{-1}B$. The density function for the distribution of the elements of $Y$ is then given by (\ref{eqn:elementjpdf}).
\end{proposition}

We delay the proof of this result until Appendix \ref{app:PY}.

\subsection{Eigenvalue jpdf and fractional linear transformation}

For a general $N \times N$ non-symmetric real matrix, we will have  $0 \leq k \leq N$ real eigenvalues, where $N$ has the same parity as $k$. From knowledge of (\ref{eqn:elementjpdf}) we can extract the eigenvalue distribution for each allowed $k$. In this task we are motivated by the work of Hough \textit{et al} \cite{HKPV2006} (see also \cite{edelman1997} and \cite{FK2009}). In particular, we work with the real Schur decomposition $Y=QR_NQ^T$, where $Q$ is real orthogonal (each column is an eigenvector of $Y$, with the restriction that the entry in the first row is positive) and
\begin{eqnarray}
\label{eqn:RN} R_N&=& \left[\begin{array}{cccccc}
\lambda_1 & ... & R_{1,k} & R_{1,k+1} & ... & R_{1,m}\\
 & \ddots & \vdots & \vdots &  & \vdots \\
 &  & \lambda_k & R_{k,k+1} & ... & R_{k,m}\\
 &  &  & z_{k+1} & ... & R_{k+1,m}\\
 & 0 &  &  & \ddots & \vdots\\
 &  &  &  &  & z_m\\
\end{array}\right],\hspace{12pt}m=(N+k)/2.
\end{eqnarray}
$R_N$ is block upper triangular: on the diagonal we have the real eigenvalues $\lambda_j$ and the $2\times2$ blocks
\begin{eqnarray}\label{17a}
\nonumber z_j &=&\left[\begin{array}{cc}
x_j & -c_j\\
b_j & x_j
\end{array}\right],\hspace{18pt}b_j,c_j>0,
\end{eqnarray}which relates to the complex eigenvalues $x_j \pm iy_j$, $y_j=\sqrt{b_jc_j}$. Note that the dimension of $R_{i,j}$ depends on its position in $R_N$:
\begin{itemize}
\item{$1\times 1$ for $i,j\leq k$,}
\item{$1\times 2$ for $i\leq k, j>k$,}
\item{$2\times 2$ for $i,j>k$.}
\end{itemize}
For this decomposition to be 1-to-1 we need the eigenvalues to be ordered and we choose
\begin{eqnarray}\label{9'}
 \lambda_1 < \cdot\cdot\cdot < \lambda_k  \qquad {\rm and}
\qquad x_{k+1}<\cdot\cdot\cdot<x_m.
\end{eqnarray}

Since we are looking to change variables from the elements of $Y$ to the eigenvalues of $Y$,
as implied by the real Schur decomposition, before proceeding we first need knowledge of the corresponding Jacobian. From  \cite{edelman1997} we know that
\begin{eqnarray}
\nonumber (dY)&=&2^{(N-k)/2}\prod_{j<p}|\lambda(R_{pp})- \lambda(R_{jj})|\\
\nonumber &&\times(d\tilde{R}_N)(Q^TdQ)\prod_{s=1}^{k}d\lambda_s \prod_{l=k+1}^{(N+k)/2} |b_l- c_l| \; dx_ldb_ldc_l,
\end{eqnarray}where $\lambda(R_{ll}) = \lambda_l$ for
$l \le k$, while $\lambda(R_{ll}) = x_l \pm iy_l$ for $l > k$, and 
$\tilde{R}_N$ is the strictly upper triangular part of $R_N$. Our interest is only in the eigenvalue dependent portion so we can immediately dispense with the dependence on $Q$ by integrating out $(Q^TdQ)$ according to
\begin{eqnarray}
\nonumber \int(Q^TdQ)=\pi^{N(N+1)/4}\prod_{j=1}^N\frac{1}{\Gamma(j/2)}
\end{eqnarray}
(see e.g.~\cite[Theorem 2.1.15]{muirhead1982}, 
with the modification of omitting the factor of $2^N$ since we have specified the columns of $Q$ to have positive first entry).

For our goal of computing the eigenvalue jpdf the objective now is to integrate over all the $\tilde{R}_{i,j}$; this procedure is similar to that in \cite{krishnapur2008} where that author was concerned with the analogous complex ensemble. The details in the present setting are contained in Appendix \ref{app:2}. According to this working (\ref{eqn:elementjpdf}) has been reduced to the distribution of $\{ \lambda_i,x_i,b_i,c_i\}$,
\begin{eqnarray}
\nonumber &&\pi^{-(N-k)/4}\Gamma((N+1)/2)^{N/2}\Gamma(N/2+1)^{N/2}\left(\frac{\Gamma((N+1)/2)}{\Gamma(N/2+1)}\right)^{k/2}\prod_{j=1}^{N-1}\frac{1}{\Gamma(j/2)^2}\\
&&\nonumber \quad  \times\prod_{s=k+1}^{(N+k)/2}\frac{1}{\det(\1+z_sz_s^T)^{N/2+1}}\prod_{s=1}^{k}\frac{1}{(1+\lambda^2_s)^{(N+1)/2}} 2^{(N-k)/2}\prod_{l=k+1}^{(N+k)/2}|b_l-c_l|\\
\label{eqn:reduced_dist}&& \quad \times \quad \prod_{j<p}|R_{pp}-R_{jj}|,
\end{eqnarray}
where use has been made of the simplification
{\small
\begin{eqnarray}
\nonumber &&\prod_{s=1}^{k-1}\pi^{(k-s)/2}\frac{\Gamma((N+1)/2)}{\Gamma((N+k-s+1)/2)}\prod_{s=0}^{(N-k)/2-1}\pi^{N-2s-2}\frac{\Gamma((N+1)/2)}{\Gamma(N-s-1/2)}\frac{\Gamma(N/2+1)}{\Gamma(N-s)}\\
\label{eqns:gammas_upside_down} \nonumber &&= \pi^{(k+N^2-2N)/4}\Gamma((N+1)/2)^{N/2}\Gamma(N/2+1)^{N/2}\left(\frac{\Gamma((N+1)/2)}{\Gamma(N/2+1)}\right)^{k/2}\prod_{s=0}^{N-1}\frac{1}{\Gamma((N+1+s)/2)}.
\end{eqnarray}
}

By writing $\epsilon_s=1+x_s^2+y_s^2$ and $\delta_s=b_s-c_s$ we see that $\det (\mathbf{1}_2+z_sz_s^T) = \epsilon_s^2+\delta_s^2$. Also, from \cite{edelman1997}, we know that
\begin{eqnarray}
\nonumber dx_s db_s dc_s&=&\frac{2y_s}{\sqrt{\delta_s^2+4y_s^2}}dx_s dy_sd\delta_s,
\end{eqnarray}
although we require $-\infty < \delta < \infty$ rather than $0< \delta < \infty$ as claimed by Edelman. So now we integrate over $\delta$
\begin{eqnarray}
\nonumber \int_{\delta=-\infty}^{\delta=\infty}\frac{|b_s-c_s|}{\det(\1+z_sz_s^T)^{N/2+1}}dx_s db_s dc_s&=&4y_s\int_{\delta=0}^{\delta=\infty}\frac{\delta \hspace{3pt}d\delta}{(\epsilon^2_s+\delta^2)^{N/2+1}\sqrt{4y^2+\delta^2}}dx_sdy_s\\
\label{eqn:delta_integ} &=&4y_s\int_{t=2y_s}^{t=\infty}\frac{dt}{(\epsilon_s^2-4y_s^2+t^2)^{N/2+1}}dx_s dy_s.
\end{eqnarray}
Substituting (\ref{eqn:delta_integ}) in (\ref{eqn:reduced_dist}) as appropriate
gives the reduced jpdf, but (\ref{eqn:delta_integ}) as written appears intractable for
further analysis. On the other hand it follows from the analysis relating to (\ref{2'}) with
$\alpha, \beta \in \mathbb R$ that when projected on to the sphere the eigenvalue density is unchanged by rotation in the $XZ$ plane, where $X,Y,Z$ are the co-ordinates after stereographic projection. This suggests that simplifications can be achieved by an appropriate mapping of the half-plane that contains the rotational symmetry of the half sphere.

We therefore introduce the fractional linear transformation (\ref{7'}) mapping the upper half-plane to the interior of the unit disk, and (\ref{14.2}) mapping the real line to the unit circle. In particular, the complicated dependence on $x_s, y_s$ in (\ref{eqn:delta_integ}) is now unravelled.

\begin{lemma}
Let $\epsilon_s = 1 + x_s^2 + y_s^2$. With the substitutions (\ref{7'}) and (\ref{14.2})
\begin{eqnarray}\label{14.3}
&&
4 y_s\int_{t=2y_s}^{t=\infty}\frac{dt}{(\epsilon_s^2-4y_s^2+t^2)^{N/2+1}}dx_s dy_s
\nonumber \\
&& \qquad =
\frac{(1-|w_s|^2)|1+w_s|^{2N-4}}{2^{2N-2}|w_s|^{N+1}}\int_{\frac{|w_s|^{-1}-|w_s|}{2}}^{\infty}\frac{dt}{\left(1+t^2\right)^{N/2+1}}du_s dv_s.
\end{eqnarray}
\end{lemma}

\textit{Proof:} Noting that
\begin{eqnarray*}
&& y_s = {1 - |w_s|^2 \over |1 + w_s|^2}, \\
&& \epsilon_s^2 - 4y_s^2 = {16 |w_s|^2 \over |1 + w_s|^4}, \\
&& dx_s dy_s = \Big | {dz_s \over dw_s} \Big |^2 du_s dv_s =
{4 \over |1 + w_s|^4} du_s dv_s,
\end{eqnarray*}
reduces the given expression to
$$
{16 \over |1 + w_s|^4} {1 - |w_s|^2 \over |1 + w_s|^2}
\int_{2(1 - |w_s|^2)/|1+w_s|^2}^\infty
{dt \over (16 (|w_s|^2/|1 + w_s|^4) + t^2 )^{N/2 + 1}} \, du_s dv_s.
$$
The RHS of (\ref{14.3}) results from this after the change of variables $t \mapsto 4|w_s| t/|1 + w_s|^2$. \hfill $\square$
\newline

For the remaining terms in (\ref{eqn:reduced_dist}), the substitutions (\ref{7'}) and (\ref{14.2}) give
{\small
\begin{eqnarray}
\nonumber \prod_{j<p}^k|\lambda_p-\lambda_j|&=&(-2i)^{k(k-1)/2}\prod_{s=1}^k\frac{(\bar{e}_s)^{(k-1)/2}}{|e_s+1|^{k-1}}\prod_{j<p}^k(e_p-e_j),\\
\nonumber \prod_{j=1}^k\prod_{s=k+1}^{(N+k)/2}|\lambda_j-z_s||\lambda_j-\bar{z}_s|&=&(-1)^{k(N-k)/2}2^{(N-k)k}\prod_{j=1}^k(\bar{e}_j)^{(N-k)/2}\left| \frac{1}{e_j+1}\right|^{(N-k)}\\
\nonumber &&\times \prod_{s=k+1}^{(N+k)/2}(\bar{w}_s)^k\left| \frac{1}{w_s+1}\right|^{2k}\prod_{j=1}^k\prod_{s=k+1}^{(N+k)/2}(w_s-e_j)\left(\frac{1}{\bar{w}_s}-e_j\right),
\end{eqnarray}
\begin{eqnarray}
\nonumber \prod_{k+1\leq a < b \leq (N+k)/2}\hspace{-24pt}&&|z_a-z_b||\bar{z}_a-\bar{z}_b|\mathop{\prod_{c,d=k+1}^{(N+k)/2}}_{c \neq d}|z_c-\bar{z}_d|=(-2)^{2\left(\frac{N-k}{2}\frac{N-k-2}{2}\right)}\prod_{j=k+1}^{(N+k)/2}(1-|w_j|^2)^{-1}\\
\nonumber &&\times\prod_{s=k+1}^{(N+k)/2}(\bar{w})^{N-k-1}\left| \frac{1}{w_s+1}\right|^{2(N-k-2)}\prod_{a < b}(w_b-w_a)\left( \frac{1}{\bar{w}_b}-\frac{1}{\bar{w}_a} \right)\prod_{c,d=k+1}^{(N+k)/2}\left( \frac{1}{\bar{w}_d}-w_c\right).
\end{eqnarray}
}

\noindent An essential feature is that, apart from the creation of some one-body terms that depend
only on the radius, the product of difference structure is conserved by the substitutions.

Combining all this, and noting
\begin{eqnarray}
\nonumber \left| \frac{1}{e_j+1}\right|^{N-1}&=&\left(\frac{1}{2}\right)^{N-1}\left(\frac{1}{\mathrm{cos}(\theta_j/2)}\right)^{N-1},
\end{eqnarray} we have the explicit form of the eigenvalue jpdf in the variables (\ref{7'}) and (\ref{14.2}).

\begin{proposition}
\label{thm:ainvb_jpdf}
Let $A,B$ be $N\times N$ matrices drawn from the real Ginibre ensemble, and let $Y=A^{-1}B$. In the variables (\ref{7'}) and (\ref{14.2}) the eigenvalue jpdf of $Y$, conditioned to have $k$ real eigenvalues ($k$ being of the same parity as $N$) is (\ref{eqn:q(y)}).
\end{proposition}

\section{Generalised partition function}
\subsection{$N$ even}
Having established the eigenvalue jpdf, we wish to find the generalised partition function from which we can calculate probabilities and correlations. For $\{e_1,\cdot\cdot\cdot,e_k\}$ corresponding to real eigenvalues and $\{w_{k+1},\cdot\cdot\cdot, w_{(N+k)/2},\bar{w}_{k+1},\cdot\cdot\cdot, \bar{w}_{(N+k)/2} \}$ corresponding to complex conjugate pairs, define the generalised partition function by 
\begin{eqnarray}
\nonumber Z_{k,(N-k)/2}[u,v]&=&\frac{1}{((N-k)/2)!}\int_0^{\theta_2}d\theta_1\int_{\theta_1}^{\theta_3}d\theta_2\cdot\cdot\cdot\int_{\theta_{k-1}}^{2\pi}d\theta_k\prod_{l=1}^ku(e_l)\\
\label{eqn:ZkN-k} &&\times\int_{\Omega}dw_{k+1}\cdot\cdot\cdot\int_{\Omega}dw_{(N+k)/2}\prod_{s=k+1}^{(N+k)/2} v(w_s)\mathcal{Q}(Y),
\end{eqnarray}
where the factor of $1/((N-k)/2)!$ comes from relaxing the ordering constraint on the complex eigenvalues, and $\Omega$ is the unit disk. Note the ordering of the angles corresponding to the real eigenvalues, in accordance with the ordering of $\{\lambda_i\}$ in (\ref{9'}).

It is at this point that parity considerations become important. For the time being, we will assume that $N$ (and consequently $k$) is even.

\begin{proposition}
\label{prop:gen_part_fn}The generalised partition function in the case where $N$ is even is 
\begin{eqnarray}
\nonumber Z_{k,(N-k)/2}[u,v]&=&\frac{(-1)^{(N/2)(N/2-1)/2}}{2^{N(N-1)/2}}\Gamma((N+1)/2)^{N/2}\Gamma(N/2+1)^{N/2}\\
\label{eqn:genpartfn} &&\times\prod_{s=1}^{N}\frac{1}{\Gamma(s/2)^2}[\zeta^{k}]\mathrm{Pf}\left[\zeta^2 \alpha_{j,l}[u,v]+\beta_{j,l}[u,v]\right],
\end{eqnarray}
with $[\zeta^k ]$ denoting the coefficient of $\zeta^k$, and where{\small
\begin{eqnarray}\label{15'} 
\nonumber \alpha_{j,k}[u,v]&=&-\frac{i}{2}\int_0^{2\pi}d\theta_1u(e_1)\tau(e_1)\int_0^{2\pi}d\theta_2 u(e_2)\tau(e_2)q_{j-1}(e_1)q_{k-1}(e_2)\mathrm{sgn}(\theta_2-\theta_1),\\
 \beta_{j,k} [u,v]&=&\int_{\Omega} dw\hspace{3pt}v(w)\tau(w)\tau\left(\frac{1} {\bar{w}} \right) \frac{1}{|w|^2}\left(q_{j-1}(w)q_{k-1}\left(\frac{1}{\bar{w}}\right) - q_{k-1}(w)q_{j-1}\left(\frac{1}{\bar{w}}\right) \right),
\end{eqnarray}
}and
\begin{eqnarray}
\label{eqn:q=p} q_{2j}(x)=p_{2j}(x),\qquad q_{2j+1}(x)=p_{N-1-2j}(x),
\end{eqnarray}
with $p_i(x)$ an arbitrary monic polynomial of degree $i$. Equivalently, with
\begin{equation}\label{ZS}
Z_N[u,v]:=  \sum_{k=0 \atop k \: {\rm even}}^N  Z_{k,(N-k)/2}[u,v],
\end{equation}
we have
\begin{eqnarray}\label{21}
Z_N[u,v] & = & \frac{(-1)^{(N/2)(N/2-1)/2}}{2^{N(N-1)/2}}\Gamma((N+1)/2)^{N/2}\Gamma(N/2+1)^{N/2}\nonumber\\
\label{eqn:genpartfn1} &&\times\prod_{s=1}^{N}\frac{1}{\Gamma(s/2)^2}\mathrm{Pf}\Big[\alpha_{j,l}[u,v]+\beta_{j,l}[u,v]\Big].
\end{eqnarray}
\end{proposition}

\textit{Proof:} With $p_i(x)$ an arbitrary monic polynomial of degree $i$,
the Vandermonde product in $\mathcal{Q}(Y)$ can be written
{\small
\begin{eqnarray}
\nonumber \Delta\left(\mathbf{e},\mathbf{w},\mathbf{\frac{1}{\bar{w}}}\right)&=&\mathrm{det}\left[\begin{array}{c}
[p_{l-1}(e_j)]_{j=1,...,k}\vspace{3pt}\\
\left[p_{l-1}(w_s)\right]_{s=k+1,...,(N+k)/2}\\
\left[p_{l-1}(1/\bar{w}_s)\right]_{s=k+1,...,(N+k)/2}
\end{array}\right]_{l=0,...,N}\\
\nonumber &=&(-1)^{(N-k)/2((N-k)/2-1)/2}\mathrm{det}\left[\begin{array}{c}
[p_{l-1}(e_j)]_{j=1,...,k}\vspace{3pt}\\
\left[\begin{array}{c}
p_{l-1}(w_s)\\
p_{l-1}(1/\bar{w}_{s})
\end{array}
\right]_{s=k+1,...,(N+k)/2}
\end{array}\right]_{l=1,...,N},
\end{eqnarray}
}

\noindent where, for the second equality, we have interlaced the rows corresponding to complex conjugate pairs; this will be convenient later. 

Next we apply the method of integration over alternate variables \cite{mehta1967,forrester?}
{\small
\begin{eqnarray}
\nonumber Z_{k,(N-k)/2}[u,v]&=&(-1)^{(N-k)/2((N-k)/2-1)/2}\frac{A_{k,N}}{(k/2)!((N-k)/2)!}\int_0^{2\pi}d\theta_2\int_{0}^{2\pi}d\theta_4\cdot\cdot\cdot\int_{0}^{2\pi}d\theta_k\\
\nonumber &&\times\int_{\Omega}dw_{k+1}\cdot\cdot\cdot\int_{\Omega}dw_{(N+k)/2} \prod_{s=k+1}^{(N+k)/2}v(w_s) \; \frac{1}{|w_s|^2}\tau(w_s)\tau\left(\frac{1}{\bar{w}_s}\right)\\
\nonumber&&\times \mathrm{det}\left[\begin{array}{c}
\left[\begin{array}{c}
\int_{0}^{\theta_{2j}}u(\theta)\tau(e)p_{l-1}(e)d\theta \\
u(\theta_{2j})\tau(e_{2j})p_{l-1}(e_{2j})\end{array}\right]_{j=1,...k/2} \vspace{6pt}\\
\left[\begin{array}{c}
p_{l-1}(w_s)\\
p_{l-1}(1/\bar{w}_s)
\end{array}
\right]_{s=k+1,...,(N+k)/2}
\end{array}\right]_{l=1,...,N}.
\end{eqnarray}
}

\noindent Re-order columns in the determinant according to
\begin{eqnarray}
\label{eqn:poly_ordering}
p_0,p_{N-1},p_2,p_{N-3},\cdot\cdot\cdot ,p_{N-2},p_1,
\end{eqnarray}
introducing a factor of $(-1)^{(N/2)(N/2-1)/2}$. For labeling purposes we define
\begin{eqnarray}
\nonumber q_{2j}(x)=p_{2j}(x),\qquad q_{2j+1}(x)=p_{N-1-2j}(x).
\end{eqnarray}
Expanding the determinant according to its definition as a signed sum over permutations,
then performing the remaining integrations gives
\begin{eqnarray*}
&& Z_{k,(N-k)/2}[u,v] = (-1)^{(N-k)/2((N-k)/2-1)/2}\frac{A_{k,N}}{(k/2)!((N-k)/2)!} \\
&& \qquad \times \sum_{P \in S_N} \epsilon(P) \prod_{l=1}^{k/2}
a_{P(2l-1),P(2l)} \prod_{l=k/2+1}^{N/2} b_{P(2l-1),P(2l)},
\end{eqnarray*}
where
\begin{eqnarray*}
a_{j,k} &=& \int_0^{2\pi} d \theta_1 \, u(\theta_1) \tau(e_1) q_{j-1}(e_1)
\int_0^{\theta_1} d \theta_2 \, u(\theta_2) \tau(e_2) q_{k-1}(e_2), \nonumber \\
b_{j,k} &=& \int_{\Omega} dw \, v(w) \tau(w) \tau\left(\frac{1}{\bar{w}}\right)\hspace{2pt}\frac{1}{\bar{w}^2}\hspace{2pt}q_{j-1}(w) q_{k-1}\left(\frac{1}{\bar{w}}\right).
\end{eqnarray*}
If we now impose the restriction $P(2l) > P(2l-1)$, ($l=1,\dots,N/2$) this can be rewritten as
\begin{eqnarray}\label{15.1}
 Z_{k,(N-k)/2}[u,v] & = & (-1)^{(N-k)/2((N-k)/2-1)/2} (2i)^{k/2} A_{k,N}
 \nonumber \\
 && \times \sum_{P \in S_N \atop P(2l) > P(2l-1)} \hspace{-12pt}\epsilon(P) \hspace{6pt}\prod_{l=1}^{k/2} \alpha_{P(2l-1),P(2l)}
 \prod_{l=k/2+1}^{N/2} \beta_{P(2l-1),P(2l)},
\end{eqnarray}
 with $\alpha_{j,k}, \beta_{j,k}$ given by (\ref{15'}). But for general $C = [c_{jk} ]_{j,k=1,\dots,N}$, $c_{jk} = - c_{kj}$ and with $N$ even,
\begin{eqnarray}
\label{def:Pf} {\rm Pf} \, C :=  \sum_{P \in S_N \atop P(2l) > P(2l-1)} \epsilon(P)
 \prod_{l=1}^{N/2} c_{P(2l-1),P(2l)}
\end{eqnarray}
allowing the sum over permutations in (\ref{15.1}) to be written in terms of a Pfaffian, and (\ref{eqn:genpartfn}) follows.
\hfill $\Box$
\newline

With $u=v=1$, $Z_{k,(N-k)/2}[u,v]$ is then the probability of finding $k$ real eigenvalues and $(N-k)/2$ complex eigenvalues, and it can be calculated using the result of Proposition \ref{prop:gen_part_fn}. The real Ginibre ensemble permits an analogous formula, which by choosing the arbitrary degree $i$ polynomials $p_i(x)$ to be the monomial $x^i$, further simplifies to involve a determinant of size $N/2 \times N/2$. For the problem at hand we can do even better, by explicitly constructing polynomials
$q_i(x)$ which for general $\zeta$ skew-diagonalise the matrix in (\ref{eqn:genpartfn}), with the $2 \times 2$ diagonal blocks of the form
\begin{eqnarray}\label{28.1}
 \left[\begin{array}{cc}
0 & \zeta^2\alpha_{2i+1,2i+2}[1,1]+\beta_{2i+1,2i+2}[1,1]\\
-(\zeta^2\alpha_{2i+1,2i+2}[1,1]+\beta_{2i+1,2i+2}[1,1]) & 0
\end{array}\right]
\end{eqnarray}
and all other entries 0.

The desired so called skew-orthogonal polynomials turn out to be quite simple, and it was knowledge of these polynomials that motivated the definition of the $\{q_i(x)\}$ in terms of the $\{p_i(x)\}$ in (\ref{eqn:q=p}). First we define the inner-products
\begin{align}
\nonumber &\langle p_j,p_l \rangle_r:= -\frac{i}{2}\int_0^{2\pi}d\theta_1\: \tau(e_1) \int_0^{2\pi} d\theta_2 \: \tau(e_2) p_j(e_1) p_l(e_2)\: \mathrm{sgn}(\theta_2-\theta_1)\\
\label{def:Sipr} &=\alpha_{j+1,l+1}[1,1],\\
\nonumber &\langle p_j,p_l \rangle_c:= \int_{\Omega} dw\; \tau(w) \tau \left(\frac{1}{\bar{w}}\right) \frac{1} {|w|^2} \left(p_j(w) p_l\left(\frac{1}{\bar{w}}\right) - p_l(w) p_j\left(\frac{1} {\bar{w}}\right) \right)\\
\label{def:Sipc} &= \beta_{j+1,l+1}[1,1],
\end{align}
and we see that the polynomials that skew-orthogonalise (\ref{def:Sipr}) and (\ref{def:Sipc}) are exactly the polynomials that yield the block-diagonal matrix (\ref{28.1}). That is, we look for polynomials that simultaneously satisfy the conditions
\begin{align}
\label{eqn:so_polys}
\begin{split}\langle p_{2j},p_{2l}\rangle_r = \langle p_{2j+1},p_{2l+1}\rangle_r =0 &,\langle p_{2j},p_{2l+1}\rangle_r =-\langle p_{2l+1},p_{2j}\rangle_r =\delta_{j,l}\: \alpha_l,\\
\langle p_{2j},p_{2l}\rangle_c = \langle p_{2j+1},p_{2l+1}\rangle_c=0 &,\langle p_{2j},p_{2l+1}\rangle_c =-\langle p_{2l+1},p_{2j}\rangle_c =\delta_{j,l}\: \beta_l,
\end{split}
\end{align}
where $\alpha_l$ and $\beta_l$ are given by (\ref{3'a}).

\begin{proposition}
\label{prop:skew_polys_even}
The skew-orthogonality conditions (\ref{eqn:so_polys}) are satisfied by the polynomials $p_j(x)=x^j$ and thus, according to (\ref{eqn:q=p}),
\begin{eqnarray}
\label{eqn:skew_polys} q_{2j}(x)=x^{2j}, \qquad q_{2j+1}(x)=x^{N-1-2j}.
\end{eqnarray}
Further, with $\alpha_j, \beta_j$ as in (\ref{3'a})
$$
\alpha_{2j+1,2j+2}[1,1] = \alpha_j, \qquad
\beta_{2j+1,2j+2}[1,1] = \beta_j,
$$
and consequently
\begin{eqnarray}
\mathrm{Pf}\left[\zeta^2\alpha_{j,l}[1,1]+\beta_{j,l}[1,1]\right]_{j,l=1,2,...,N}&=&\prod_{l=0}^{N/2-1}(\zeta^2\alpha_l+\beta_l).
\end{eqnarray}
\end{proposition}

\textit{Proof:} The skew-symmetry property $\alpha_{j,l}[1,1]= -\alpha_{l,j}[1,1]$, $\beta_{j,l}[1,1]= -\beta_{l,j}[1,1]$ can be checked by observation, so to establish the result we must show that both $\alpha_{j,l}[1,1]$ and $\beta_{j,l}[1,1]$ are non-zero only for $j=2t+1,l=2t+2$ in which case they have the evaluations stated.

From (\ref{15'}), we have
\begin{align}
\nonumber \alpha_{j+1,l+1}[1,1]&=c \: \frac{i}{2}\int_0^{2\pi}d\theta_1\: e^{i\theta_1 (\tilde{j} - (N-1)/2)}\int_{\theta_1}^{2\pi}d\theta_2\:  e^{i\theta_2(\tilde{l}- (N-1)/2}\\
\nonumber &-c \: \frac{i}{2}\int_0^{2\pi}d\theta_1\: e^{i\theta_1 (\tilde{j} - (N-1)/2)}\int_0^{\theta_1}d\theta_2\:  e^{i\theta_2(\tilde{l}- (N-1)/2},
\end{align}
where $c$ is a constant factor and $\tilde{j}=2j$ or $\tilde{j}=N-1-2j$ for $j$ even or odd respectively. Performing the inner integrals over $\theta_2$, using the fact that $\tilde{j}\neq (N-1)/2$ since $j$ is an integer and $N$ is even, we find
\begin{align}
\label{eqn:sopsa1} \alpha_{j+1,l+1}[1,1]&=c\: \frac{2}{2\tilde{l}-N+1} \int_0^{2\pi}d\theta_1\: e^{i\theta_1 (\tilde{j} +\tilde{l} - N+1)},
\end{align}
which is non-zero only in the case that $\tilde{l}=N-1-\tilde{j}$, or $j=2t+1,l=2t+2$ (for $\alpha_{j+1,l+1}[1,1]$ positive). The evaluation of (\ref{eqn:sopsa1}) in this case is straightforward. To obtain the conditions on $s$ and $t$ where $\beta_{s,t}[1,1]\neq 0$ we repeat the procedure used above by writing out $\tau(w),\tau(\bar{w}^{-1}),q_{2s},q_{2t+1}$. The fact that $s=2j+1$ and $t=2j+2$ for a non-zero evaluation is then immediate.

It remains to show that $\beta_{2j+1,2j+2}[1,1]= \beta_{j}$, which requires knowledge of a non-standard form of the beta integral. After converting to polar co-ordinates, setting $c:=|w|^2$ and integrating by parts, one obtains
\begin{eqnarray}
\nonumber \beta_{2j+1,2j+2}&=&-\frac{2\pi^{3/2}}{N-1-4j}\frac{\Gamma((N+1)/2)}{\Gamma(N/2+1)}+\frac{2^{N+1}\pi}{N-1-4j}\int_0^1\frac{c^{2j}+c^{N-2j-1}}{(1+c)^{N+1}}dc.
\end{eqnarray}
According to \cite[Equation 3.216 (1)]{GR94}, for general $a,b$ such that Re$\, b>0$,
Re$\, (a-b) > 0$,
$$
\int_0^1(t^{b-1} + t^{a-b-1})(1 + t)^{-a} \, dt = \frac{\Gamma(b)\Gamma(a-b)}{\Gamma(a)},
$$
and thus, with $b = y$, $a-b = x$, is a non-standard form of the beta integral
$$
\int_0^1 t^{x-1}(1-t)^{y-1} \, dt = \frac{\Gamma(x)\Gamma(y)}{\Gamma(x+y)}.
$$
The stated formula for $\beta_{2s+1,2t+2}$ now follows.
\hfill $\square$
\newline

This simplification allows us to give a simple product form for the generating function of the probabilities $\{p_{N,k}\}$, and to proceed to specify statistical properties of the corresponding distribution.

\begin{proposition}
\label{prop:gen_fn}
Let $N$ be even.
The generating function for $p_{N,k}$, the probability of finding $k$ real eigenvalues, as specified by
\begin{eqnarray}
\label{def:ZNxi} Z_N(\xi)&:=& \sum_{k=0 \atop k \: {\rm even}}^N \xi^k p_{N,k} \: = \: \sum_{k=0}^{N/2}\xi^{2k} Z_{2k,(N-2k)/2}[1,1],
\end{eqnarray}
has the evaluation (\ref{eqn:Z_N}).
\end{proposition}

Using (\ref{def:ZNxi}) we calculate the expected number of real eigenvalues and the variance.

\begin{corollary}\label{3.5}
With $N$ even the expected number of real eigenvalues is
\begin{eqnarray}\label{29}
E_N=\left. \frac{\partial}{\partial \xi}Z_N(\xi) \right|_{\xi=1}= \sum_{l=0}^{N/2-1} \frac{2\hspace{1pt} \alpha_l}{\alpha_l+\beta_l},
\end{eqnarray}
which has evaluation given by (\ref{eqn:eks_result}). The variance in the number of real eigenvalues is
\begin{eqnarray}\label{30}
\nonumber \sigma_N^2 &=& {\partial^2 \over \partial \xi^2} Z_N(\xi) \Big |_{\xi = 1} + E_N - E_N^2 \: = \: 2E_N - 4 \sum_{l=0}^{N/2 - 1} {\alpha_l^2 \over (\alpha_l + \beta_l)^2},
\end{eqnarray}
which has evaluation (\ref{3.1a}).
\end{corollary}

\noindent
\textit{Proof:} The second equalities follow from the first and
(\ref{eqn:Z_N}), while to obtain (\ref{eqn:eks_result}) and (\ref{3.1a}) use has been made of the summations
$$
\sum_{j=0}^{N/2 - 1} \Big ( {N - 1 \atop 2j} \Big ) = 2^{N-2}, \qquad
\sum_{j=0}^{N/2 - 1} \Big ( {N - 1 \atop 2j} \Big )^2 = {2^{2N-3} \; \Gamma (N-1/2) \over \sqrt{\pi} \; \Gamma(N)}.
$$
{}\hfill $\square$
\newline

As noted in the Introduction, the result (\ref{29}) was first derived by Edelman \textit{et al.}~\cite{eks1994} using ideas from integral geometry. A corollary, also noted in \cite{eks1994}, is that for $N \to \infty$
$$
E_N \: \sim \: \sqrt{\pi N \over 2} \bigg ( 1 - {1 \over 4N} + {1 \over 32 N^2} +
{5 \over 128 N^3} - {21 \over 2048 N^4} + {\rm O} \Big ( {1 \over N^5} \Big ) \bigg ).
$$
It was also remarked in the Introduction that to leading order (\ref{3.1a}) implies the variance is related to the mean by $\sigma_N^2 \sim (2 - \sqrt{2} )E_N$, which coincidentally (?) is the same asymptotic relation as found in the case of the real Ginibre ensemble \cite{forrester and nagao2007}.

The explicit form of the generating function given in Proposition \ref{prop:gen_fn} allows for the computation of the large $N$ limiting form of the probability density of the scaled number of real eigenvalues.

\begin{proposition}\label{pAB}
Let $\sigma_N^2$ and $E_N$ be as in Corollary \ref{3.5}, and let $[ \cdot ]$ denote the integer part.
We have
\begin{eqnarray}
\nonumber \lim_{N \to \infty} \: {\rm sup}_{x \in (-\infty,\infty)}
\Big | \sigma_N p_{N,[\sigma_N x + E_N]} - {1 \over \sqrt{2 \pi}} e^{- x^2/2} \Big | = 0.
\end{eqnarray}
\end{proposition}

\noindent
\textit{Proof:} For a given $n$, let $\{p_n(k)\}_{k=0,1,\dots,n}$ be a sequence such that
$$
P_N(x) = \sum_{k=0}^n p_n(k) x^k
$$
has the properties that the zeros of $P_N(x)$ are all on the real axis, and $P_N(1)=1$.
Let
$$
\mu_n = \sum_{k=0}^n k p_n(k), \qquad \sigma_n^2 = \sum_{k=0}^n k^2 p_n(k) - \mu_n^2
$$
and suppose $\sigma_n \to \infty$ as $n \to \infty$. A local limit theorem due to Bender 
\cite{Be73} gives
\begin{eqnarray}
\lim_{n \to \infty} {\rm sup}_{x \in (-\infty,\infty)}
\Big | \sigma_n p_n([\sigma_n x + \mu_n]) - {1 \over \sqrt{2 \pi}} e^{-x^2/2} \Big | = 0.
\end{eqnarray}
Application of this general theorem to $Z_N(\xi)$, with $\xi^2 = x$, gives the stated result.
\hfill $\square$
\newline

\begin{figure}
\label{fig:pnk}
\begin{center}
\includegraphics[scale=0.6,trim=0 0mm 80mm 540, clip=true]{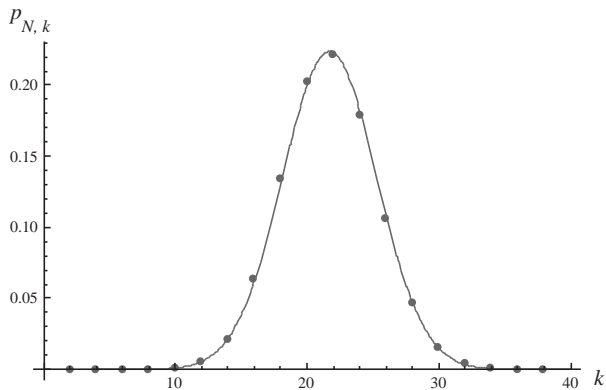}
\end{center}
\caption{A plot of $p_{300,k}$, that is, the probability of finding $k$ real eigenvalues from a $300\times 300$ real matrix $Y=A^{-1}B$, where $A,B$ are real matrices with iid Gaussian elements. These values were calculated using (\ref{eqn:Z_N}). The solid line is the Gaussian curve from Proposition \ref{pAB} with a normalising factor of $2$ since $N$ and $k$ must be of the same parity.}
\end{figure}

In Figure \ref{fig:pnk} we have used (\ref{3a}) to calculate the value of $p_{N,k}$ for $N=300,k={2,4,...,38}$ and overlaid it with the Gaussian curve given by Proposition \ref{pAB}; the agreement is quite clear.

\subsection{$N$ odd}
Here we will deal with the case of $N$ odd. The parity of $N$ has no bearing on (\ref{eqn:q(y)}), nor on (\ref{eqn:ZkN-k}). However, the proof of Proposition \ref{prop:gen_part_fn} uses the method of integration over alternate variables, which crucially depends on the evenness of $N$ as it pairs the real eigenvalues. In the case of $N$ odd, one of the real eigenvalues cannot be paired; it is the treatment of this eigenvalue that distinguishes the even and odd cases.

\begin{proposition} With $\alpha_{j,l}[u,v]$ and $\beta_{j,l}[u,v]$ as in (\ref{15'}), the generalised partition function in the case of $N$ odd is
\label{prop:gen_part_fn_odd}
{\small
\begin{eqnarray}
\nonumber &&Z_{k,(N-k)/2}[u,v] =\frac{(-1)^{(N-1)/4((N-1)/2)-1)}}{2^{N(N-1)/2}}\Gamma((N+1)/2)^{N/2}\Gamma(N/2+1)^{N/2}\\
\label{eqn:gen_fn_odd} &&\times\prod_{s=1}^{N}\frac{1}{\Gamma(s/2)^2}[\zeta^{k-1}]\mathrm{Pf}\left[\begin{array}{cc}
\left[\zeta^2\alpha_{i,j}[u,v]+\beta_{i,j}[u,v]\right]_{i,j=1,...,N} & \left[\nu_i[u]\right]_{i=1,...,N}\\
 \left[-\nu_j[u]\right]_{j=1,...,N} & 0\\
\end{array}\right],
\end{eqnarray}
}

\noindent where, with $e:=e^{i\theta}$ (recall (\ref{14.2})),
\begin{eqnarray}
\label{def:nu} \nu_l[u]:=\frac{1}{\sqrt{2}}\int_0^{2\pi}u(\theta)\tau(e)q_{l-1}(e) d\theta,
\end{eqnarray}
and
\begin{eqnarray}
\nonumber &&\left.
\begin{array}{l}
q_{2j}=p_{2j},\\
q_{2j+1}=p_{N-1-2j},
\end{array}
\right\}0 \leq 2j<(N-1)/2,\\
\nonumber \\
\nonumber &&\left.
\begin{array}{l}
q_{2j}=p_{2j+1},\\
q_{2j+1}=p_{N-1-(2j+1)},
\end{array}
\right\}(N-1)/2 \leq 2j<N-1,\\
\nonumber \\
\label{eqn:odd_polys} &&\left.
\begin{array}{l}
q_{N-1}=p_{(N-1)/2}.
\end{array}
\right.
\end{eqnarray}
The analogous definition to (\ref{ZS}) for $Z_N[u,v]$ where $N$ is odd is
\begin{eqnarray}
\label{eqn:gen_fn_odd1} Z_N[u,v]:=  \sum_{k=1 \atop k \: {\rm odd}}^N  Z_{k,N-k/2}[u,v],
\end{eqnarray}
and
\begin{eqnarray}
\nonumber Z_N[u,v] &=&\frac{(-1)^{(N-1)/4((N-1)/2)-1)}}{2^{N(N-1)/2}}\Gamma((N+1)/2)^{N/2}\Gamma(N/2+1)^{N/2}\\
\nonumber &&\times\prod_{s=1}^{N}\frac{1}{\Gamma(s/2)^2}\mathrm{Pf}\left[\begin{array}{cc}
\left[\alpha_{i,j}[u,v]+\beta_{i,j}[u,v]\right]_{i,j=1,...,N} & \left[\nu_i[u]\right]_{i=1,...,N}\\
 \left[-\nu_j[u]\right]_{j=1,...,N} & 0\\
\end{array}\right].
\end{eqnarray}
\end{proposition}

\textit{Proof:} Similar to the even case, again with arbitrary monic polynomials $\{p_i(x)\}$, we write the Vandermonde product of $\mathcal{Q}(Y)$ as
{\small
\begin{align}
\nonumber \Delta\left(\mathbf{e},\mathbf{w},\mathbf{\frac{1}{\bar{w}}}\right)&=\mathrm{det}\left[\begin{array}{c}
[p_{l-1}(e_j)]_{j=1,...,k-1}\vspace{3pt}\\
\left[p_{l-1}(w_s)\right]_{s=k+1,...,(N+k)/2}\\
\left[p_{l-1}(1/\bar{w}_s)\right]_{s=k+1,...,(N+k)/2}
\end{array}\right]_{l=0,...,N}\\
\label{eqn:odd_vand} &=(-1)^{(N-k)/2((N-k)/2-1)/2}\mathrm{det}\left[\begin{array}{c}
[p_{l-1}(e_j)]_{j=1,...,k}\vspace{3pt}\\
\left[\begin{array}{c}
p_{l-1}(w_s)\\
p_{l-1}(1/\bar{w}_{s})
\end{array}
\right]_{s=k+1,...,(N+k)/2}\\
\left[ p_{l-1}(e_k)\right]
\end{array}\right]_{l=1,...,N},
\end{align}
}

\noindent where we have moved the row corresponding to the $k$th real eigenvalue to the bottom of the matrix. This always involves an even number of transpositions so no overall factor is required. It is more convenient converting this matrix to Pfaffian form than the equivalent matrix where the $k$th row is not moved. This row corresponds to the single unpaired real eigenvalue that must exist in any odd-sized real matrix, a fact which is guaranteed by $N$ and $k$ being of the same parity.

Now we substitute (\ref{eqn:odd_vand}) in (\ref{eqn:ZkN-k}) and apply integration over alternate variables, as in Proposition \ref{prop:gen_part_fn}, to find
{\small
\begin{align}
\nonumber &Z_{k,(N-k)/2}[u,v]=(-1)^{(N-k)/2((N-k)/2-1)/2}\frac{A_{k,N}}{((k-1)/2)!((N-k)/2)!}\\
\nonumber & \times \int_0^{2\pi}d\theta_2\int_{0}^{2\pi}d\theta_4\cdot\cdot\cdot\int_{0}^{2\pi}d\theta_{k-1} \int_{\Omega}dw_{k+1} \cdot\cdot\cdot\int_{\Omega} dw_{(N+k)/2} \prod_{s=k+1}^{(N+k)/2} v(w_s) \; \frac{1}{|w_s|^2}\tau(w_s)\tau\left(\frac{1}{\bar{w}_s}\right)\\
\nonumber &\times \mathrm{det}\left[\begin{array}{c}
\left[\begin{array}{c}
\int_{0}^{\theta_{2j}}u(\theta)\tau(e)p_{l-1}(e)d\theta \\
u(\theta_{2j})\tau(e_{2j})p_{l-1}(e_{2j})\end{array}\right]_{j=1,...(k-1)/2} \vspace{6pt}\\
\left[\begin{array}{c}
p_{l-1}(w_s)\\
p_{l-1}(1/\bar{w}_s)
\end{array}
\right]_{s=k+1,...,(N+k)/2}\\
\\
\left[ \int_{0}^{2\pi}u(\theta)\tau(e)p_{l-1}(e)\hspace{2pt}d\theta \right]
\end{array}\right]_{l=1,...,N}.
\end{align}
}

We need to reorder the columns of the determinant in a similar way to that of (\ref{eqn:poly_ordering}), although with the key difference of shifting the middle column to the end. As in the even case, this is to assist in the use of skew-orthogonal polynomials. The re-ordering becomes
\begin{eqnarray}
\nonumber &&p_0,p_{N-1},p_2,p_{N-3},\cdot\cdot\cdot ,p_{(N-1)/2-\epsilon_{1,2}},p_{(N-1)/2+\epsilon_{1,2}},\\
\label{eqn:poly_order_odd} &&p_{(N-1)/2+\epsilon_{2,1}},p_{(N-1)/2-\epsilon_{2,1}},\cdot\cdot\cdot p_{N-4},p_{3},p_{N-2},p_{1},p_{(N-1)/2},
\end{eqnarray}
where
\begin{eqnarray}
\nonumber \epsilon_{1,2}=\left\{
\begin{array}{cc}
1 & \mbox{for $(N-1)/2$ even,}\\
2 & \mbox{for $(N-1)/2$ odd,}
\end{array}
\right.\\
\nonumber \epsilon_{2,1}=\left\{
\begin{array}{cc}
2 & \mbox{for $(N-1)/2$ even,}\\
1 & \mbox{for $(N-1)/2$ odd.}
\end{array}
\right.
\end{eqnarray} 
This introduces a factor of $(-1)^{(N-1)/2+(N-1)/2((N-1)/2-1)/2}$. Also, for $N$ odd, the factors of $(-1)$ in $A_{k,n}$ can be re-written by noting
\begin{eqnarray}
\nonumber (-1)^{(N-k)k/2-k(k-1)/4}=(-1)^{(N-1)/2-(k-1)/4}.
\end{eqnarray}
This gives us an overall factor of
\begin{eqnarray}
\nonumber &&(-1)^{(N-k)/2((N-k)/2-1)/2}\times (-1)^{(N-1)/2+(N-1)/2((N-1)/2-1)/2}\times A_{k,N}\\
\nonumber &&= \frac{(-1)^{(N-1)/4((N-1)/2)-1)-(k-1)/4}}{2^{(N(N-1)+k)/2}}\Gamma((N+1)/2)^{N/2}\Gamma(N/2+1)^{N/2}\prod_{s=1}^{N}\frac{1}{\Gamma(s/2)^2}.
\end{eqnarray}
Now we again expand the determinant as a signed sum over permutations and impose the restriction $P(2l)>P(2l-1)$. This gives us the odd analogue of (\ref{15.1})
{\small
\begin{eqnarray}
\nonumber Z_{k,(N-k)/2}[u,v]  =  \frac{(-1)^{(N-1)/4((N-1)/2)-1)}}{2^{N(N-1)/2}}\Gamma((N+1)/2)^{N/2}\Gamma(N/2+1)^{N/2}\prod_{s=1}^{N}\frac{1}{\Gamma(s/2)^2}\\
\nonumber \\
\nonumber  \quad \times \sum_{P \in S_N \atop P(2l) > P(2l-1)} \hspace{-12pt}\epsilon(P) \; \nu_{P(N),N+1} \prod_{l=1}^{(k-1)/2} \alpha_{P(2l-1),P(2l)} \prod_{l=(k+1)/2}^{(N-1)/2} \beta_{P(2l-1),P(2l)}
\end{eqnarray}
}

\noindent where $\nu_{P(N)}[u]:=\nu_{P(N),N+1}$ is given by (\ref{def:nu}). Using the Pfaffian definition (\ref{def:Pf}), (\ref{eqn:gen_fn_odd}) now follows.
\hfill $\square$
\newline

To simplify the calculation of the Pfaffian for the even case in (\ref{eqn:genpartfn}) we used the skew-orthogonal polynomials of Proposition \ref{prop:skew_polys_even}. We would like to find equivalent polynomials for the odd case, that is polynomials that will reduce the Pfaffian in (\ref{eqn:gen_fn_odd}) to the block diagonal form
\begin{eqnarray}
\label{eqn:skew_mat_odd} \left[
\begin{array}{cccc}
B_1 & 0 & \cdots & 0\\
0 & \ddots &  & \vdots\\
\vdots & & B_{(N-1)/2}&0\\
0 & \cdots & 0 &\left[\begin{array}{cc}
0 & \nu_N\\
-\nu_N& 0
\end{array}\right]
\end{array}
\right],
\end{eqnarray}
where the $B_i$ are the $2\times 2$ blocks given by (\ref{28.1}). However, the best we can do here is to obtain the structure
\begin{align}
\label{eqn:modskodd} \left[\begin{array}{cc}
\left[\begin{array}{ccc}
B_1 &  & 0\\
 & \ddots & \\
0 & & B_{(N-1)/2}
\end{array}\right] & \left[\begin{array}{cc}
G_{1,N} & 0\\
\vdots & \vdots\\
G_{N,N} & 0
\end{array}\right]\\
\\
\left[\begin{array}{ccc}
-G_{1,N} & \cdot\cdot\cdot & -G_{N,N}\\
0 & \cdot\cdot\cdot & 0
\end{array}\right] & \left[\begin{array}{cc}
0 & \nu_N\\
-\nu_N& 0
\end{array}\right]
\end{array}
\right],
\end{align}
where $G_{s,N}:=\zeta\alpha_{s,N}[1,1]+\beta_{s,N}[1,1]$. The structure (\ref{eqn:modskodd}) will be sufficient for our ends since the Pfaffians of (\ref{eqn:skew_mat_odd}) and (\ref{eqn:modskodd}) are equal, which can be seen by applying the Laplace expansion for Pfaffians.

By comparing  (\ref{eqn:modskodd}) with (\ref{28.1}) we see that we are looking to skew-orthogonalise the same inner products (\ref{def:Sipr}) and (\ref{def:Sipc}) as in the even case, however those inner products are dependent on $N$. Also, the column reordering (\ref{eqn:poly_order_odd}) means while these polynomials are still monomials, compared to the even case the labeling is more complicated, since there was the additional movement of the middle column to the end. The first half of the polynomials are the same as the even case, while the second half are modified by $j\rightarrow j+1/2$. The middle polynomial must be singled out for special treatment. For these reasons, the specification of the skew-orthogonal polynomials for $N$ odd is different (and more complicated) from that for $N$ even.

\begin{proposition} \label{prop:odd_polys}
The matrix
\begin{eqnarray}
\nonumber \left[\begin{array}{cc}
\left[\zeta^2\alpha_{i,j}[1,1]+\beta_{i,j}[1,1]\right]_{i,j=1,...,N} & \left[\nu_i[1]\right]_{i=1,...,N}\\
 \left[-\nu_j[1]\right]_{j=1,...,N} & 0\\
\end{array}\right]
\end{eqnarray}
evaluates to the modified block diagonal form of (\ref{eqn:modskodd}) using the polynomials $p_j(x)=x^j$ ($j\neq(N-1)/2$), and thus according to (\ref{eqn:odd_polys})
\begin{eqnarray}
\nonumber &&\left.
\begin{array}{l}
q_{2j}(x)=x^{2j},\\
q_{2j+1}(x)=x^{N-1-2j},
\end{array}
\right\}0 \leq 2j<(N-1)/2,\\
\nonumber \\
\nonumber &&\left.
\begin{array}{l}
q_{2j}(x)=x^{2j+1},\\
q_{2j+1}(x)=x^{N-1-(2j+1)},
\end{array}
\right\}(N-1)/2 \leq 2j<N-1,
\end{eqnarray}
provided the degree $(N-1)/2$ polynomial $q_{N-1}(x)$ is chosen as
{\small
\begin{eqnarray}
\nonumber q_{N-1}(x)&=&x^{(N-1)/2}+\sum_{j=0}^{ (N-1)/2-1}\left(\frac{\langle q_{2j+1},x^{(N-1)/2}\rangle_r + \langle q_{2j+1},x^{(N-1)/2}\rangle_c } {\alpha_{2j+1,2j+2}[1,1]+ \beta_{2j+1,2j+2}[1,1]}\; q_{2j}(x)\right.\\
\nonumber &&\left. -\; \frac{\langle q_{2j},x^{(N-1)/2}\rangle_r + \langle q_{2j},x^{(N-1)/2}\rangle_c}{\alpha_{2j+1,2j+2}[1,1]+\beta_{2j+1,2j+2}[1,1]}\; q_{2j+1}(x)\right).
\end{eqnarray}
}
\noindent With these polynomials
\begin{eqnarray}
\nonumber &&\left.
\begin{array}{l}
\alpha_{2j+1,2j+2}[1,1]=\alpha_j,\\
\beta_{2j+1,2j+2}[1,1]=\beta_j,
\end{array}
\right\}0 \leq 2j<(N-1)/2,\\
\nonumber \\
\nonumber &&\left.
\begin{array}{l}
\alpha_{2j+1,2j+2}[1,1]=\alpha_{j+1/2},\\
\beta_{2j+1,2j+2}[1,1]=\beta_{j+1/2},
\end{array}
\right\}(N-1)/2 \leq 2j<N-1,\\
\nonumber \\
\nonumber &&\left.
\begin{array}{l}
\nonumber \alpha_{s,N}[1,1]+\beta_{s,N}[1,1]=\alpha_{N,s}[1,1]+\beta_{N,s}[1,1]=0, \qquad s \leq N,
\end{array}
\right.
\\
\nonumber &&\left.
\begin{array}{l}
\nonumber v_N:=\nu_N[1]=\pi \sqrt{\frac{\Gamma((N+1)/2)}{\Gamma(N/2+1)}},
\end{array}
\right.
\end{eqnarray}
where $\alpha_j,\beta_j$ are as in (\ref{3'a}) and
\begin{eqnarray}
\nonumber \alpha_{j+1/2}&=&\frac{2\pi}{N-3-4j}\frac{\Gamma((N+1)/2)}{\Gamma(N/2+1)},\\
\nonumber &&\\
\nonumber \beta_{j+1/2}&=&\frac{2\sqrt{\pi}}{N-3-4j}\left( 2^N\frac{\Gamma(2j+2)\Gamma(N-2j-1)}{\Gamma(N+1)}-\sqrt{\pi}\frac{\Gamma((N+1)/2)}{\Gamma(N/2+1)}\right).
\end{eqnarray}
So in particular
\begin{eqnarray}
\nonumber \mathrm{Pf}\left[\begin{array}{cc}
\left[\zeta^2\alpha_{i,j}[1,1]+\beta_{i,j}[1,1]\right]_{i,j=1,...,N} & \left[\nu_i[1]\right]_{i=1,...,N}\\
\left[-\nu_j[1]\right]_{j=1,...,N} & 0\\
\end{array}\right]\\
\label{eqn:oddPfgpf} =\nu_N\prod_{l=0}^{\lceil (N-1)/4\rceil-1}(\zeta^2\alpha_l+\beta_l)\prod_{l=\lceil (N-1)/4\rceil}^{(N-1)/2-1}(\zeta^2\alpha_{l+1/2}+\beta_{l+1/2}),
\end{eqnarray}
where $\lceil x\rceil$ is the ceiling function on $x$.
\end{proposition}

\textit{Proof:} For $0\leq 2j<(N-1)/2$ we have the result by Proposition \ref{prop:skew_polys_even} and replacing $j\mapsto j+1/2$ we have result for $(N-1)/2\leq 2j < N-1$. By the construction of $q_{N-1}$, we see that $\alpha_{s,N}[1,1]+\beta_{s,N}[1,1]=0$ for $1\leq s\leq N$. All that remains is to show that the row and column of $\nu_l$ obey the skew-orthogonality condition.

Writing out the $\nu_l[1]$ in full, the fact that it is non-zero only for $l=N$ is clear, that is, only when $l=N$ does the angular dependence cancel from the integral. In which case the evaluation is straightforward.

Recall from above that the Pfaffian of the modified block diagonal structure (\ref{eqn:modskodd}) is equal to that of the Pfaffian of (\ref{eqn:skew_mat_odd}) and so we have the evaluation in (\ref{eqn:oddPfgpf}).
\hfill $\Box$
\newline

The odd analogue of (\ref{eqn:Z_N}) can now be given.

\begin{proposition}
\label{prop:gen_fn_xi_odd}
In the case of $N$ odd, the generating function for $p_{N,k}$ is
\begin{eqnarray}
\nonumber Z_N(\xi):=\sum_{k=1 \atop k \: {\rm odd}}^N \xi^k p_{N,k} \: = \: \sum_{l=0}^{(N-1)/2}\xi^{2l+1} Z_{2l+1,(N-2l-1)/2}[1,1]\
\end{eqnarray}
and has evaluation
\begin{eqnarray}\label{51a}
\nonumber Z_N(\xi)&=&\frac{(-1)^{(N-1)/4((N-1)/2)-1)}}{2^{N(N-1)/2}}\Gamma((N+1)/2)^{N/2}\Gamma(N/2+1)^{N/2}\prod_{s=1}^{N}\frac{1}{\Gamma(s/2)^2}\\
 &&\times\hspace{3pt}\xi\nu_N\prod_{l=0}^{\lceil (N-1)/4\rceil-1}(\xi^2\alpha_l+\beta_l)\prod_{l=\lceil (N-1)/4\rceil}^{(N-1)/2-1}(\xi^2\alpha_{l+1/2}+\beta_{l+1/2}).
\end{eqnarray}
\end{proposition}

From Proposition \ref{prop:gen_fn_xi_odd} we can calculate the expected number of real eigenvalues in the case of $N$ odd, which we know from \cite{eks1994} is given by (\ref{eqn:eks_result}) independent of the parity of $N$. Similarly, we can check that the formula (\ref{3.1a}) for the variance also holds independent of the parity of $N$.

\begin{corollary}
\label{prop:odd_EN_var}
For $N$ odd, the expected number of real eigenvalues of $Y$ can be written
\begin{eqnarray}
\nonumber E_N =1+\sum_{l=0}^{\lceil (N-1)/4\rceil-1}\frac{2\hspace{2pt}\alpha_l}{\alpha_l+\beta_l}+\sum_{l=\lceil (N-1)/4\rceil}^{(N-1)/2-1}\frac{2\hspace{2pt}\alpha_{l+1/2}}{\alpha_{l+1/2}+\beta_{l+1/2}},
\end{eqnarray}
which has evaluation (\ref{eqn:eks_result}).

The variance for $N$ odd is
{\small
\begin{eqnarray}
\hspace{-40pt}\nonumber \sigma_N^2&=&2(E_N-1)-\sum_{l=0}^{\lceil (N-1)/4\rceil-1}\frac{4\hspace{2pt}\alpha_l^2}{(\alpha_l+\beta_l)^2}+\sum_{l=\lceil (N-1)/4\rceil}^{(N-1)/2-1}\frac{4\hspace{2pt}\alpha_{l+1/2}^2}{(\alpha_{l+1/2}+\beta_{l+1/2})^2},
\end{eqnarray}
}
which has evaluation (\ref{3.1a}).
\end{corollary}

\textit{Proof:} The formulae in terms of $\{ \alpha_l,\beta_l \}$ follow from
Proposition \ref{prop:gen_fn_xi_odd} and for the expressions for $E_N$, $\sigma_N^2$
in terms of $Z_N(\xi)$ recall (\ref{3.1}) and (\ref{6.1}). 
For the summations we use the identity 
\begin{eqnarray}
\nonumber \sum_{l=0}^{\lceil (N-1)/4\rceil-1}{N-1 \choose 2l}^p+\sum_{l=\lceil (N-1)/4\rceil}^{(N-1)/2-1}{N-1 \choose 2l+1}^p&=&\sum_{l=0}^{(N-1)/2-1}{N-1 \choose l}^p
\end{eqnarray}
for integer $p$ and for both $(N-1)/4\in\mathbb{Z}$ and $(N-1)/4\in\mathbb{Z}+1/2$.
\hfill $\Box$
\newline

\begin{table}
\begin{center}
\begin{tabular}{|c|c|c|c|} \hline
&$\mathrm{Exact}\hspace{6pt}p_{N,k}$&Decimal $p_{N,k}$&Simulated $p_{N,k}$\\
\hline
$p_{2,2}$ \T &$\frac{1}{4}\pi$&$0.785398$&$0.78691$\\
$p_{2,0}$ \T &$1-\frac{1}{4}\pi$&$0.214602$&$0.21309$\\
\hline
$p_{3,3}$ &$\frac{1}{2}$ \T &$0.5$&$0.50051$\\
$p_{3,1}$ &$\frac{1}{2}$ \T &$0.5$&$0.49949$\\
\hline
$p_{4,4}$&$\frac{27}{1024}\pi^2$ \T &$0.260234$&$0.25705$\\
$p_{4,2}$&$\frac{3}{8}\pi-\frac{27}{512}\pi^2$ \T &$0.65763$&$0.66053$\\
$p_{4,0}$&$1-\frac{3}{8}\pi+\frac{27}{1024}\pi^2$ \T &$0.0821365$&$0.08242$\\
\hline
$p_{5,5}$&$\frac{1}{9}$ \T &$0.111111$&$0.11167$\\
$p_{5,3}$&$\frac{11}{18}$ \T &$0.611111$&$0.60969$\\
$p_{5,1}$&$\frac{5}{18}$ \T &$0.277778$&$0.27864$\\
\hline
$p_{6,6}$&$\frac{84375}{67108864}\pi^3$ \T &$0.0389837$&$0.03898$\\
$p_{6,4}$&$\frac{14625}{262144}\pi^2-\frac{253125}{67108864}\pi^3$ \T &$0.433673$&$0.43216$\\
$p_{6,2}$&$\frac{15}{32}\pi-\frac{14625}{131072}\pi^2+\frac{253125}{67108864}\pi^3$ \T &$0.488323$&$0.48873$\\
$p_{6,0}$&$1-\frac{15}{32}\pi+\frac{14625}{262144}\pi^2-\frac{84375}{67108864}\pi^3$ \T &$0.0390194$&$0.04013$\\
\hline
$p_{7,7}$&$\frac{9}{800}$ \T &$0.01125$&$0.01178$\\
$p_{7,5}$&$\frac{39}{160}$ \T &$0.24375$&$0.24244$\\
$p_{7,3}$&$\frac{463}{800}$ \T &$0.57875$&$0.57933$\\
$p_{7,1}$&$\frac{133}{800}$ \T &$0.16625$&$0.16645$\\
\hline
\end{tabular}
\end{center}
\caption{Calculations of $p_{N,k}$, the probability of finding $k$ real eigenvalues from an $N\times N$ matrix $Y=A^{-1}B$. The second column is the analytic calculation, the third column is the analytic calculation in decimal. These are compared to the right column, which contains the results of a numerical simulation of 100,000 matrices.}\label{table:pnk}
\end{table}

The values of $p_{N,k}$ for $N=2,...,7$, calculated using Propositions \ref{prop:gen_fn} and \ref{prop:gen_fn_xi_odd}, are listed in Table \ref{table:pnk}, along with the results of a simulation of 100,000 matrices. A remarkable fact can be immediately seen in the table: the probabilities for even $N$ are polynomials in $\pi$ of degree $N/2$, while for odd $N$ they are rational numbers. The key difference is that $(N+1)/2$ and $N/2+1$ alternate as integers and half integers, depending on whether $N$ is even or odd. These values introduce factors of $\sqrt{\pi}$ through the gamma functions.

\begin{proposition}\label{p310}
Let $p_{N,k}$ be the probability of finding $k$ real eigenvalues in a matrix $Y=A^{-1}B$, where $A,B$ are Gaussian real. Then for $N$ even, $p_{N,k}$ is a polynomial in $\pi$ of degree $N/2$. For $N$ odd, $p_{N,k}$ is a rational number. 
\end{proposition}

\textit{Proof:} For $N$ even $\alpha_l$ and the second term in $\beta_l$ (with $u=v=1$) both
yield factors of $\pi^{3/2}$. The pre-factor in (\ref{eqn:genpartfn}) yields $\pi^{-N/4}$. Combining these two we find the highest power of $\pi$ is $N/2$. Noting that the first term in $\beta_l$ has a factor of $\pi^{1/2}$ and expanding the product in (\ref{eqn:genpartfn})  gives lower order terms in $\pi$.

For the odd case, the pre-factor in (\ref{51a}) gives $\pi^{-N/4-1/2}$. Then by noting that 
$(\xi^2 \alpha_l + \beta_l)$ and $(\xi^2 \alpha_{l+1/2} + \beta_{l+1/2})$ both give factors of
$\pi^{1/2}$ and $\nu_N$ gives $\pi^{3/4}$ we see that
 the end result is a rational number.
\hfill $\Box$

\section{Correlation functions}
We would like to make use of knowledge of the Pfaffian form of the generating function (\ref{21}), and the skew-orthogonal polynomials (\ref{eqn:skew_polys}), to compute the correlation functions $\rho_{(k_1,k_2)}$. The latter specifies the probability density for $k_1$ eigenvalues occurring at specific points on the unit circle, and $k_2$ eigenvalues occurring at specific points in the unit disk. Note that there is no conditioning on the number of real eigenvalues. The probability density is normalised so that $\rho_{(k_1+1,k_2)}/\rho_{(k_1,k_2)}$ corresponds to the density of eigenvalues at a specific point on the unit circle, given the location of the $k_1$ eigenvalues on the unit circle already specified, and the $k_2$ eigenvalues in the disk already specified. It can be calculated in terms of the summed up generalised partition function (\ref{ZS}) by functional differentiation,
\begin{equation}\label{ZS1}
\rho_{(k_1,k_2)}(\mathbf{e},\mathbf{w})
= {1 \over Z_N[u,v]} {\delta^{k_1+k_2} \over \delta u(e_1) \cdots \delta u(e_{k_1})
\delta v(w_1) \cdots \delta v(w_{k_2}) } Z_N[u,v] \Big |_{u=v=1}.
\end{equation}

To compute (\ref{ZS1}) from the formula (\ref{21}) for $Z_N[u,v]$, we draw on established theory relating to calculation of $\rho_{(k_1,k_2)}$ for the real Ginibre ensemble. For $N$ even, the summed up generalised grand partition function for the real Ginibre ensemble is proportional to \cite{Si06,forrester and nagao2007}
\begin{equation}\label{S1}
{\rm Pf} \, [\tilde{\alpha}_{j,k} + \tilde{\beta}_{j,k} ]_{j,k=1,\dots,N},
\end{equation}
where, for arbitrary monic polynomials $p_i(x)$ of degree $i$,
\begin{eqnarray}\label{12.E}
\tilde{\alpha}_{j,k} & = & \int_{-\infty}^\infty dx \, u(x)
\int_{-\infty}^\infty dy \, u(y) \,
e^{-(x^2 + y^2)/2} p_{j-1}(x) p_{k-1}(y) {\rm sgn} \, (y - x), \nonumber \\
\tilde{\beta}_{j,k} & = & 2 i \int_{{\mathbb R}_+^2} dx dy \, v(x,y) e^{y^2 - x^2}
{\rm erfc} (\sqrt{2} y) \nonumber \\
&& \times \Big ( p_{j-1}(x+iy) p_{k-1}(x-iy) - p_{k-1}(x+iy) p_{j-1}(x-iy) \Big ).
\end{eqnarray}
Comparing this to (\ref{21}) and (\ref{15'}) we see that upon the identifications
$p_k(x+iy) \leftrightarrow q_k(w)$, 
$p_k(x-iy) \leftrightarrow q_k(1/\bar{w})$, the two expressions are structurally identical.
In the case of (\ref{S1}), with $\{p_k(x)\}$ chosen to have skew-orthogonality properties 
analogous to (\ref{eqn:so_polys}), a $(k_1 + k_2) \times (k_1 + k_2)$ Pfaffian formula for
$\rho_{(k_1,k_2)}$ has been deduced which makes use of the structural properties
exhibited by (\ref{12.E}) (see \cite{BS08}); this can be adapted to the
present problem, allowing us to deduce the following explicit evaluation.

\begin{theorem}
\label{thm:correlns}
Let $D$ denote the unit disk in the complex plane, and let $\partial D$ denote its boundary, the
unit circle in the complex plane. For $N$ even
\begin{eqnarray}
\label{eqn:correlns} \rho_{(k_1,k_2)}(\mathbf{e},\mathbf{w})=\mathrm{Pf}\left[\begin{array}{cc}
K_N(e_i,e_j) & K_N(e_i,w_m)\\
K_N(w_l,e_j) & K_N(w_l,w_m)\\
\end{array}\right],\qquad e_i\in \partial D, \hspace{3pt} w_i \in D,
\end{eqnarray}
\begin{eqnarray}
\label{eqn:kernel} K_N(s,t)=\left[\begin{array}{cc}
D(s,t) & S(s,t)\\
-S(t,s) & I(s,t)\\
\end{array}\right],
\end{eqnarray}
with$$\mathbf{e}=\{ e_1,...,e_{k_1} \} \hspace{6pt},\hspace{6pt} \mathbf{w}=\{ w_1,...,w_{k_2} \}$$
where 
\begin{eqnarray}
\nonumber D(x_i,x_j)&=&\sum_{l=0}^{\frac{N}{2}-1}\frac{1}{r_l}\Bigl[a_{2l}(x_i)a_{2l+1}(x_j)-a_{2l+1}(x_i)a_{2l}(x_j)\Bigr],\\
\nonumber S(x_i,x_j)&=&\sum_{l=0}^{\frac{N}{2}-1}\frac{1}{r_l}\Bigl[a_{2l}(x_i)b_{2l+1}(x_j)-a_{2l+1}(x_i)b_{2l}(x_j)\Bigr],\\
\nonumber I(x_i,x_j)&=&\sum_{l=0}^{\frac{N}{2}-1}\frac{1}{r_l}\Bigl[b_{2l}(x_i)b_{2l+1}(x_j)-b_{2l+1}(x_i)b_{2l}(x_j)\Bigr]+\epsilon(x_i,x_j),
\end{eqnarray}
and
\begin{eqnarray}
\nonumber a_j(x) &=& 
\left\{ 
\begin{array}{ll}
|x|^{-1}\tau(x)\hspace{2pt}q_j(x),  & x\in  D, \\
\sqrt{-i/2}\hspace{2pt}\tau(x) q_j(x),  & x\in \partial D,\\
\end{array}
\right.\\
\nonumber\\
\nonumber b_j(x) &=& 
\left\{ 
\begin{array}{ll}
|x|^{-1}\tau(\bar{x}^{-1})\hspace{2pt}q_j(\bar{x}^{-1}),  & x\in D,\\
\sqrt{-i/2}\int_{0}^{2\pi}\tau(e^{i\theta})q_j(e^{i\theta})\mathrm{sgn}(\theta-\mathrm{arg}(x))d\theta,  & x\in \partial D,\\
\end{array}
\right.\\
\nonumber\\
\nonumber \epsilon(x_i,x_j) &=& 
\left\{ 
\begin{array}{ll}
\mathrm{sgn}(\mathrm{arg}(x_i)-\mathrm{arg}(x_j)),  & x_i,x_j\in \partial D,\\
0,  & \mathrm{otherwise},\\
\end{array}
\right.\\
\nonumber r_l&=&\alpha_l+\beta_l,
\end{eqnarray}
and the polynomials $\{q_{i}(x)\}$ are as in (\ref{eqn:skew_polys}).
\end{theorem}

Due to the polynomials of Proposition \ref{prop:odd_polys} being skew-orthogonal in the sense of (\ref{eqn:modskodd}) we can adapt the results of \cite{Si08} to the present problem to yield the correlations for $N$ odd.

\begin{theorem}
For $N$ odd, the correlations obey the same Pfaffian form as in (\ref{eqn:correlns}) above, with the kernel structure as in (\ref{eqn:kernel}). The kernel elements are given by
\begin{eqnarray}
\nonumber &&D(x_i,x_j)=\sum_{l=0}^{\lceil (N-1)/4\rceil -1}\frac{1}{r_l}\Bigl[a_{2l}(x_i)a_{2l+1}(x_j)-a_{2l+1}(x_i)a_{2l}(x_j)\Bigr]\\
\nonumber &&+\sum_{l=\lceil (N-1)/4\rceil}^{(N-1)/2 -1}\frac{1}{r_{l+1/2}}\Bigl[a_{2l}(x_i)a_{2l+1}(x_j)-a_{2l+1}(x_i)a_{2l}(x_j)\Bigr],\\
\nonumber &&S(x_i,x_j)=\sum_{l=0}^{\lceil (N-1)/4\rceil -1}\frac{1}{r_l}\Bigl[a_{2l}(x_i)b_{2l+1}(x_j)-a_{2l+1}(x_i)b_{2l}(x_j)\Bigr]\\
\nonumber &&+\sum_{l=\lceil (N-1)/4\rceil}^{(N-1)/2 -1}\frac{1}{r_{l+1/2}}\Bigl[a_{2l}(x_i)b_{2l+1}(x_j)-a_{2l+1}(x_i)b_{2l}(x_j)\Bigr]+\kappa(x_i,x_j),\\
\nonumber &&I(x_i,x_j)=\sum_{l=0}^{\lceil (N-1)/4\rceil -1}\frac{1}{r_l}\Bigl[b_{2l}(x_i)b_{2l+1}(x_j)-b_{2l+1}(x_i)b_{2l}(x_j)\Bigr]\\
\nonumber &&+\sum_{l=\lceil (N-1)/4\rceil}^{(N-1)/2 -1}\frac{1}{r_{l+1/2}}\Bigl[b_{2l}(x_i)b_{2l+1}(x_j)-b_{2l+1}(x_i)b_{2l}(x_j)\Bigr]+\epsilon(x_i,x_j)+\sigma(x_i,x_j),
\end{eqnarray}
where $a,b$ and $\epsilon(x_i,x_j)$ are as in Theorem \ref{thm:correlns},
\begin{eqnarray}
\nonumber \kappa(x_i,x_j) &=& 
\left\{ 
\begin{array}{ll}
\frac{\tau(x_i)}{\nu_N}q_{N-1}(x_i),  & x_j\in \partial D,\\
0,  & \mathrm{otherwise},\\
\end{array}
\right.\\
\nonumber\\
\nonumber \sigma(x_i,x_j) &=& 
\left\{ 
\begin{array}{ll}
\frac{1}{\nu_N}(b_{N-1}(x_i)-b_{N-1}(x_j)),  & x_i,x_j\in \partial D,\\
-\frac{1}{\nu_N}b_{N-1}(x_j), &x_i\in\partial D,x_j\in D,\\
\frac{1}{\nu_N}b_{N-1}(x_i), &x_i\in D,x_j\in\partial D,\\
0,  & \mathrm{otherwise},\\
\end{array}
\right.\\
\nonumber r_{l+1/2}&=&\alpha_{l+1/2}+\beta_{l+1/2},
\end{eqnarray}
and the polynomials $\{q_{i}(x)\}$ are as in Proposition \ref{prop:odd_polys}.
\end{theorem}

\subsection{Kernel element evaluations}

Clearly, the correlations in (\ref{eqn:correlns}) are completely determined by the kernel $K_N(s,t)$ of (\ref{eqn:kernel}). The elements of the kernel satisfy the following relations
\begin{eqnarray}
\nonumber I_{r,r}(e_1,e_2)&=&\int_{\theta_1}^{\theta_2}S_{r,r}(e,e_2) d\theta+\mathrm{sgn}(\theta_1-\theta_2),\\
\nonumber D_{r,r}(e_1,e_2)&=&\frac{\partial}{\partial \theta_2} S_{r,r}(e_1,e_2),\\
\nonumber I_{r,c}(x,w)&=&\int_x^wS_{r,c}(z,w)dz,\\
\nonumber D_{r,c}(e_1,w)&=&\frac{1}{|w|^2}S_{r,c}(e_1,\bar{w}^{-1}),\\
\nonumber I_{c,c}(w_1,w_2)&=&\frac{1}{|w_1|^2}S_{c,c}(\bar{w}_1^{-1},w_2),\\
\label{eqn:kernelrelations} D_{c,c}(w_1,w_2)&=&\frac{1}{|w_2|^2}S_{c,c}(w_1,\bar{w}_2^{-1}),
\end{eqnarray}
where the subscripts denote the real or complex nature of the two arguments. It is clear from these relationships that the various $S(s,t)$ determine the nature of the kernel block (\ref{eqn:kernel}). These $S(s,t)$ can be written in a summed-up form which is independent of the parity of $N$.

\begin{proposition} \label{prop:summedupS}
The elements of the correlation kernel (\ref{eqn:kernel}) $S_{r,r}(s,t)$, $S_{r,c}(s,t)$, $S_{c,r}(s,t)$ and $S_{c,c}(s,t)$, corresponding to real-real, real-complex, complex-real and complex-complex eigenvalue pairs respectively, can be evaluated as
\begin{eqnarray}
\nonumber S_{r,r}(e_1,e_2)&=&\frac{\Gamma((N+1)/2)}{2\sqrt{\pi}\Gamma(N/2)}\mathrm{cos}\left(\frac{\theta_2-\theta_1}{2}\right)^{N-1},
\end{eqnarray}
\begin{eqnarray}
\nonumber S_{c,r}(w,e_1)&=&\left(\frac{-i}{\sqrt{\pi}}\right)^{1/2}\frac{1}{r_w}\frac{iN}{2^N\sqrt{\pi}}\sqrt{\frac{\Gamma((N+1)/2)}{\Gamma(N/2+1)}}\left[ \int_{\frac{r_w^{-1}-r_w}{2}}^{\infty}\frac{dt}{(1+t^2)^{N/2+1}}\right]^{1/2}\\
\nonumber &&\times\left( \frac{e^{-i(\theta_w-\theta_1)/2}}{r_w^{1/2}}+\frac{e^{i(\theta_w-\theta_1)/2}}{r_w^{-1/2}} \right)^{N-1},\\
\nonumber S_{r,c}(e_1,w)&=&\left( \frac{-i}{\sqrt{\pi}} \right)^{1/2}\frac{1}{r_w}\frac{N(N-1)}{2^{N+2}\sqrt{\pi}}\left[ \int_{\frac{r_w^{-1}-r_w}{2}}^{\infty}\frac{dt}{(1+t^2)^{N/2+1}}\right]^{1/2}\sqrt{\frac{\Gamma((N+1)/2)}{\Gamma(N/2+1)}}\\
\nonumber &&\times \left( \frac{e^{-i(\theta_1-\theta_w)/2}}{r_w^{1/2}} +\frac{e^{i(\theta_1-\theta_w)/2}}{r_w^{-1/2}}\right)^{N-2}\left( \frac{e^{-i(\theta_1-\theta_w)/2}}{r_w^{1/2}} -\frac{e^{i(\theta_1-\theta_w)/2}}{r_w^{-1/2}}\right),\\
\nonumber S_{c,c}(w,z)&=&\frac{N(N-1)}{2^{N+1}\pi r_w r_z}\left[\int_{\frac{r_w^{-1}-r_w}{2}}^{\infty}\frac{dt}{\left(1+t^2\right)^{N/2+1}}\right]^{1/2}\left[\int_{\frac{r_z^{-1}-r_z}{2}}^{\infty}\frac{dt}{\left(1+t^2\right)^{N/2+1}}\right]^{1/2}\\
\nonumber &&\times \left( \frac{e^{i(\theta_z-\theta_w)/2}}{(r_wr_z)^{1/2}}+\frac{e^{-i(\theta_z-\theta_w)/2}}{(r_wr_z)^{-1/2}} \right)^{N-2} \left( \frac{e^{i(\theta_z-\theta_w)/2}}{(r_wr_z)^{1/2}}-\frac{e^{-i(\theta_z-\theta_w)/2}}{(r_wr_z)^{-1/2}} \right),
\end{eqnarray}
where $w,z:=r_we^{i\theta_w},r_ze^{i\theta_z}$.
\end{proposition}

\textit{Proof:} Using the binomial theorem, the identity
\begin{eqnarray}
\nonumber \frac{1}{2}\left(\frac{d}{dx}(1+x)^{2n-1}+\frac{d}{dx}(1-x)^{2n-1}\right)=\sum_{p=0}^{n-1}2p {2n-1 \choose 2p}x^{2p-1}
\end{eqnarray}
and the results of Proposition \ref{prop:skew_polys_even} (for the even case) and Proposition \ref{prop:odd_polys} (for the odd case) the respective sums can be performed.
\hfill $\Box$
\newline
Note that $\int_{0}^{2\pi}S_{r,r}(e_1,e_1) \, d \theta = E_N$, providing a further derivation of (\ref{eqn:eks_result}).

The simplest cases of Theorem \ref{thm:correlns} are $(k_1,k_2) = (1,0)$ and $(k_1,k_2) = (0,1)$. Since these correspond to the real and complex densities respectively, we write $\rho_{(1,0)}(e) = \rho_{(1)}^{\rm r}(\theta)$ and $\rho_{(0,1)}(w) = \rho_{(1)}^{\rm c}(w)$. According to Theorem \ref{thm:correlns}, $\rho_{(1)}^{\rm r}(\theta) = S_{r,r}(e,e)$ and $\rho_{(1)}^{\rm c}(w) = S_{c,c}(w,w)$, and we read off from Proposition \ref{prop:summedupS} the evaluations (\ref{eqn:rhor}) and (\ref{eqn:rhoc}) respectively. Recalling (\ref{eqn:eks_result}), (\ref{eqn:rhor}) has the large $N$ form
$$
\rho_{(1)}^{\rm r}(\theta)  \sim {1 \over 2\pi } \sqrt{{\pi N \over 2 }},
$$
while integration by parts of (\ref{eqn:rhoc}) shows
$$
\rho_{(1)}^{\rm c}(w) \sim {(N-1) \over \pi} {1 \over (1 + r^2)^2} - {N - 1 \over N - 2}
{1 \over \pi} {1 \over (1 - r^2)^2} + \rm{O}\Big ( {1 \over N} \Big )
$$
valid for $r \in [0, 1 - {\rm O}(1/\sqrt{N})]$. To leading order in $N$ the eigenvalue density is 
therefore equal to
\begin{equation}\label{df}
{N \over \pi} {1 \over (1 + r^2)^2}
\end{equation}
for all $r \in [0,1]$. This, projected stereographically onto the half sphere, gives a uniform distribution. This $1/N$ convergence should be contrasted with the exponential convergence in the case of the polynomials (\ref{eqn:rand_polys}) (see \cite{Mc09}).

\subsection{Averages over characteristic polynomials}

As emphasised in \cite{FK06,APS08} there is a large class of eigenvalue jpdfs such that the eigenvalue density is given in terms of an average over the corresponding characteristic polynomials. This is true of the one-point function (density) for the complex eigenvalues, with $N\to N+2$ (for convenience), in the present generalised eigenvalue problem for which the jpdf is given by (\ref{eqn:q(y)}). Thus write
\begin{eqnarray}
\label{eqn:charpoly} C_N(z)=\prod_{j=1}^k(z-e_j)\prod_{s=k+1}^{(N+k)/2}(z-w_s)(z-1/\bar{w}_s)
\end{eqnarray}
for the characteristic polynomial in the $N\times N$ case of $Y=A^{-1}B$ conditioned to have $k$ real eigenvalues, with eigenvalues transformed according to (\ref{7'}) and (\ref{14.2}). Letting
\begin{align}
\nonumber \tilde{G}_{N}:=\frac{(-1)^{(N/2)(N/2-1)/2}}{2^{N(N-1)/2}} \Gamma((N+1)/2)^{N/2} \Gamma(N/2+1)^{N/2} \prod_{s=1}^{N}\frac{1}{\Gamma(s/2)^2},
\end{align}
which is the pre-factor in (\ref{3a}), then it follows from (\ref{eqn:q(y)}) and the definition of the density that
\begin{eqnarray}
\label{eqn:charpolydensity} \rho_{(1)}^{(N+2 , c)}(z)=\frac{\tilde{G}_{N+2}}{\tilde{G}_N} \frac{1}{|z|^2}\tau(z)\tau(1/\bar{z})(1/\bar{z}-z)\langle C_N(z)C_N(1/\bar{z})\rangle,
\end{eqnarray}
where the superscript $N+2$ denotes the number of eigenvalues in the system. Of course we can therefore read off from (\ref{eqn:rhoc}) the exact form of the average in (\ref{eqn:charpolydensity}). Moreover, in keeping with the development in \cite{APS08}, we can use our integration methods to compute the more general average $\langle C_N(z_1)C_N(1/\bar{z}_2) \rangle$ which we expect to be closely related to $S_{c,c}(z_1,z_2)$, in accordance with known results from the real, complex and real quaternion Ginibre ensemble \cite{Ka02,AV03,AB07,APS08,SW08}.

Note that we will also introduce a superscript on $\alpha$ and $\beta$ to indicate the size of system that they relate to, that is $\alpha_{j,l}^{(t)}$ has $j,l=1,...,t$, and $\alpha_{s}^{(t)}$ are the corresponding normalisations.

\begin{proposition}
With the characteristic polynomial $C_N$ given by (\ref{eqn:charpoly}) and $\langle\cdot\rangle$ an average with respect to (\ref{eqn:q(y)}) summed over $k$, one has
\begin{align}
\nonumber \langle C_N(z_1)C_N(1/\bar{z}_2) \rangle&=\frac{\tilde{G}_{N}}{\tilde{G}_{N+2}} (1/\bar{z}_2-z_1)^{-1}\\
\nonumber &\times \sum_{s=0}^{N/2}\frac{1}{\alpha_s^{(N+2)}+ \beta_s^{(N+2)}} \left( z_1^{2s}(1/\bar{z}_2)^{2s+1}-z_1^{2s+1}(1/\bar{z}_2)^{2s}\right)\\
\label{eqn:charpolyD} &=\frac{\tilde{G}_{N}}{\tilde{G}_{N+2}} (1/\bar{z}_2 -z_1)^{-1} \left(\frac{ \tau(z_1)}{|z_1|} \frac{\tau(1/\bar{z}_2)} {|z_2|}\right)^{-1} \left.D_{c,c}(z_1,1/\bar{z}_2) \right|_{N\to N+2},
\end{align}
where it is assumed N is even. Furthermore
\begin{eqnarray}
\nonumber \left.S_{c,c}(z_1,z_2)\right|_{N \to N+2} &=& \frac{\tilde{G}_{N+2}}{\tilde{G}_N} \frac{\tau(z_1)}{|z_1|}\frac{\tau(1/\bar{z}_2)}{|z_2|}(1/\bar{z}_2-z_1)\langle C_N(z_1)C_N(1/\bar{z}_2) \rangle
\end{eqnarray}
from which we reclaim (\ref{eqn:charpolydensity}).
\end{proposition}

\textit{Proof:} From (\ref{eqn:q(y)}) we see that
\begin{eqnarray}
\nonumber  C_N(z_1)C_N(1/\bar{z}_2)\mathcal{Q}(Y) =A_{k,N}\prod_{j=1}^k\tau(e_j)\prod_{s=k+1}^{(N+k)/2}\frac{1}{|w_s|^2}\tau(w_s)\tau\left(\frac{1}{\bar{w}_s}\right)\\
\nonumber \times(1/\bar{z}_2-z_1)^{-1}\Delta\left(\mathbf{e},\mathbf{w},\mathbf{\frac{1}{\bar{w}}},z_1,\frac{1}{\bar{z}_2}\right).
\end{eqnarray}
Integrating over $\mathbf{e}$ and $\mathbf{w}$ gives
\begin{align}
\nonumber \langle C_N(z_1)C_N(1/\bar{z}_2) \rangle\: \rule[-8pt]{0.5pt}{18pt}_{\: k \: {\rm fixed}} &= \tilde{G}_N (1/\bar{z}_2-z_1)^{-1}\\
\nonumber &\times [\varkappa^{N/2}][\zeta^{k}]\mathrm{Pf}\left[\varkappa(\zeta^2 \alpha_{j,l}^{(N+2)}+\beta_{j,l}^{(N+2)})+\gamma_{j,l}^{(N+2)}\right],
\end{align}
where $\gamma_{j,l}^{(t)}=q_{j-1}(z_1)q_{l-1}(1/\bar{z}_2)-q_{l-1}(z_1)q_{j-1}(1/\bar{z}_2)$ $(j,l=1,...,t)$, and the parameters $u,v$ in $\alpha[u,v]$ and $\beta[u,v]$ are taken to be one. Summing over $k$ leads to
\begin{align}
\nonumber \langle C_N(z_1)C_N(1/\bar{z}_2)\rangle&:=  \sum_{k=0 \atop k \: {\rm even}}^N  \langle C_N(z_1)C_N(1/\bar{z}_2)\rangle\: \rule[-8pt]{0.5pt}{18pt}_{\: k \: {\rm fixed}}\\
\nonumber &=\tilde{G}_N (1/\bar{z}_2-z_1)^{-1}[\varkappa^{N/2}]\mathrm{Pf} \left[\varkappa( \alpha_{j,l}^{(N+2)} +\beta_{j,l}^{(N+2)} )+ \gamma_{j,l}^{(N+2)} \right].
\end{align}
Using the skew-orthogonal polynomials (\ref{eqn:skew_polys}), we find
\begin{eqnarray}
\nonumber &&[\varkappa^{N/2}]\mathrm{Pf}\left[\varkappa(\alpha_{j,l}^{(N+2)}+ \beta_{j,l}^{(N+2)} )+\gamma_{j,l}^{(N+2)}\right]\\
\nonumber &&\qquad =\sum_{s=0}^{N/2}\gamma_{2s+1,2s+2}\prod_{j=0 \atop j\neq s}^{N/2}(\alpha_{2j+1,2j+2}^{(N+2)} +\beta_{2j+1,2j+2}^{(N+2)}).
\end{eqnarray}
The evaluation now follows upon recalling the form of $Z_{N+2}(1)=1=\tilde{G}_{N+2} \prod_{l=0}^{N/2} (\alpha_l^{(N+2)}+\beta_l^{(N+2)})$ implied by Proposition \ref{prop:gen_fn}.

The expression for $\left.S_{c,c}(z_1,z_2)\right|_{N\to N+2}$ is a simple manipulation of (\ref{eqn:charpolyD}).
\hfill $\Box$

\subsection{Scaled limit}

Before  implementing the fractional linear transformations (\ref{7'}) and (\ref{14.2}), we have from (\ref{eqn:rho}) that the density of real eigenvalues near the origin is proportional to $E_N$, and thus $\sqrt{N}$. A scaled limit involves changing the variables so that this density, and that of the complex eigenvalues, becomes of order unity. Such a limiting procedure is of interest because the resulting correlations are expected to be the same as for the generalised eigenvalue problem with entries chosen from general zero mean and finite variance distributions, and furthermore the same as for the eigenvalues of the real Ginibre ensemble, scaled near the origin.

In the complex case, an analogy between the eigenvalue jpdf of the generalised eigenvalue problem and the Boltzmann factor for the two-dimensional one-component plasma on a sphere \cite{caillol81}, together with the analogy between the eigenvalue jpdf for the Ginibre matrices and the two-dimensional one-component plasma in the plane \cite{AJ80} allow this latter point to be anticipated from a Coulomb gas perspective.

The limiting correlation of the eigenvalues in the vicinity of the origin for the real Ginibre ensemble have recently been computed in \cite[Corollary 9]{BS08} (see also \cite{forrester and nagao2007,SW08}).

\begin{proposition}
\label{prop:BS_limiting_kernels}
For random $N\times N$ real Ginibre matrices the correlations for the eigenvalues in the vicinity of the origin are, in the limit $N\to\infty$, given by
\begin{eqnarray}
\label{eqn:BSlargeN} \rho_{k_1,k_2}(\mathbf{x},\mathbf{z})=\mathrm{Pf}\left[\begin{array}{cc}
[K_{r,r}(x_j,x_l)]_{j,l=1,...,k_1} & [K_{r,c}(x_j,z_l)]_{j=1,...,k_1 \atop l=1,...,k_2}\\
\left(\left[-K_{r,c}(x_j,z_l)\right]_{j=1,...,k_1 \atop l=1,...,k_2}\right)^T & [K_{c,c}(z_j,z_l)]_{j,l=1,...,k_2}
\end{array}\right],
\end{eqnarray}
with
\begin{eqnarray}
\nonumber K_{r,r}(x,y)&=&\left[\begin{array}{cc}
\frac{1}{\sqrt{2\pi}}(y-x)e^{-(x-y)^2/2}&\frac{1}{\sqrt{2\pi}}e^{-(x-y)^2/2}\\
-\frac{1}{\sqrt{2\pi}}e^{-(x-y)^2/2}&\frac{1}{2}\mathrm{sgn}(x-y)\mathrm{erfc}\left(\frac{|x-y|}{\sqrt{2}}\right)
\end{array}\right],\\
\nonumber \\
\nonumber K_{r,c}(x,w)&=&\frac{1}{\sqrt{2\pi}}\sqrt{\mathrm{erfc}(\sqrt{2}\mathrm{Im}(z))}\\
\nonumber &&\times \left[\begin{array}{cc}
(w-x)e^{-(x-w)^2/2}&i(\bar{w}-x)e^{-(x-\bar{w})^2/2}\\
-e^{-(x-w)^2/2}&-ie^{-(x-\bar{w})^2/2}
\end{array}\right],\\
\nonumber \\
\nonumber K_{c,c}(w,z)&=&\frac{1}{\sqrt{2\pi}}\sqrt{\mathrm{erfc}(\sqrt{2}\mathrm{Im}(w))\mathrm{erfc}(\sqrt{2}\mathrm{Im}(z))}\\
\nonumber && \times \left[\begin{array}{cc}
(z-w)e^{-(w-z)^2/2}&i(\bar{z}-w)e^{-(w-\bar{z})^2/2}\\
i(z-\bar{w})e^{-(\bar{w}-z)^2/2}&-(\bar{z}-\bar{w})e^{-(\bar{w}-\bar{z})^2/2}
\end{array}\right].
\end{eqnarray}
\end{proposition}

In the present problem, with our use of the transformed variables (\ref{7'}) and (\ref{14.2}), the original origin has been mapped to $(1,0)$. We must choose scaled co-ordinates so that in the vicinity of this point the real and complex eigenvalues have a density of order unity. For the real eigenvalues, from the knowledge that their expected value is of order $\sqrt{N}$ and that they are uniform on the unit circle, with $e_j:=e^{ix_j}$, we scale
\begin{eqnarray}
\label{eqn:largeNreal} x_j\mapsto \frac{2X_j}{\sqrt{N}}.
\end{eqnarray}
For the complex eigenvalues, which total of order $N$ in the unit disk, an order one density will result by writing
\begin{eqnarray}
\label{eqn:largeNcomplex} w_j\mapsto 1+\frac{2i}{\sqrt{N}}W_j.
\end{eqnarray}
Note that the real and imaginary parts have been interchanged to match the geometry of the problem in the Ginibre ensemble, that is so the eigenvalues are again distributed in the upper half-plane, including the real line. The factors of $2$ in (\ref{eqn:largeNreal}) and (\ref{eqn:largeNcomplex}) are included so an exact correspondence with the results of Proposition \ref{prop:BS_limiting_kernels} can be obtained.

Since $S_{r,r}(x,x)$ is interpreted as a density, the normalised quantity is $S_{r,r}(x,x)dx$. It follows then that in the more general case we must look at the scaled limit of $S_{r,r}(x,y)\sqrt{dxdy}$ and $S_{c,c}(w_1,w_2)\sqrt{d^2w_1d^2w_2}$. For $S_{r,c}(x,w)$ and $S_{c,r}(w,x)$ we require that the product $S_{r,c}(x,w)S_{c,r}(w,x)dxd^2w$ has a well defined limit. From (\ref{eqn:largeNreal}) and (\ref{eqn:largeNcomplex}) we see
\begin{eqnarray}
\nonumber \sqrt{dxdy}&\mapsto&\frac{2}{\sqrt{N}}\sqrt{dXdY},\\
\nonumber dxd^2w&\mapsto&\left( \frac{4}{N}\right)^{3/2}dXd^2W,\\
\nonumber \sqrt{d^2w_1d^2w_2}&\mapsto&\frac{4}{N}\sqrt{d^2W_1d^2W_2}.
\end{eqnarray}
With this change of variables the large $N$ form of the correlation kernel for the spherical ensemble matches that of the Ginibre ensemble.

\begin{proposition}
\label{prop:scaled_limit}
Recall $K_N (s,t)$ from (\ref{eqn:kernel}). Replacing $x_j$ and $w_j$ according to (\ref{eqn:largeNreal}) and (\ref{eqn:largeNcomplex}) then taking $N\to\infty$ gives
\begin{eqnarray}
\nonumber \frac{2}{\sqrt{N}}K_N(e_i,e_j)&\sim& K_{r,r}(X_i,X_j),\\
\nonumber \frac{2^{5/2}}{N}K_N(e_i,w_j)&\sim& K_{r,c}(X_i,W_j),\\
\nonumber \frac{\sqrt{2}}{\sqrt{N}}K_N(w_i,e_j)&\sim& -\left(K_{r,c}(W_i,X_j)\right)^{T},\\
\nonumber \frac{4}{N}K_N(w_i,w_j)&\sim& K_{c,c}(W_i,W_j).
\end{eqnarray}
\end{proposition}

\textit{Proof:} From the explicit functional forms of Proposition \ref{prop:summedupS}, we see that elementary limits suffice. For example, changing variables $t\mapsto 2t/\sqrt{N}$ shows
\begin{eqnarray}
\nonumber \int_{\frac{r_w^{-1}-r_w}{2}}^{\infty}\frac{dt}{(1+t^2)^{N/2+1}}\sim\sqrt{\frac{\pi}{2N}}\: \mathrm{erfc}(\sqrt{2}\mathrm{Im}W).
\end{eqnarray} 
Combining such calculations we obtain
\begin{eqnarray}
\nonumber \frac{2}{\sqrt{N}}S_{r,r}(e_i,e_j)&\sim&\frac{1}{\sqrt{2\pi}}e^{-(X_i-X_j)^2/2},\\
\nonumber \frac{2^{5/2}}{N}S_{r,c}(e_i,w_j)&\sim&\frac{\sqrt{-i}}{\sqrt{2\pi}}\sqrt{\mathrm{erfc}(\sqrt{2}\mathrm{Im}W_j)}\: e^{-(X_i-\overline{W}_j)^2/2}i(\overline{W}_j-X_i),\\
\nonumber \frac{\sqrt{2}}{\sqrt{N}}S_{c,r}(w_i,e_j)&\sim&\frac{\sqrt{-i}}{\sqrt{2\pi}}\sqrt{\mathrm{erfc}(\sqrt{2}\mathrm{Im}W_i)}\: ie^{-(W_i-X_j)^2/2},\\
\nonumber \frac{4}{N}S_{c,c}(w_i,w_j)&\sim&\frac{1}{\sqrt{2\pi}}\sqrt{\mathrm{erfc}(\sqrt{2}\mathrm{Im}W_i)}\sqrt{\mathrm{erfc}(\sqrt{2}\mathrm{Im}W_j)}\\
\nonumber &&\times i(\overline{W}_j-W_i)e^{-(W_i-\overline{W}_j)^2/2},
\end{eqnarray}
which is in agreement with the off-diagonal entries on the RHS of the present proposition, as implied by Proposition \ref{prop:BS_limiting_kernels} (when one recalls that $S_{r,c}$ and $S_{c,r}$ never appear individually; only as the product $S_{r,c}S_{c,r}$).

Recalling the inter-relationships (\ref{eqn:kernelrelations}), the other kernel elements $D$ and $I$ can be obtained from $S$, giving the diagonal entries required by Proposition \ref{prop:BS_limiting_kernels}.\hfill $\Box$

\subsection{Sum rules}
With $\Delta$ as in (\ref{eqn:q(y)}) we have
\begin{eqnarray*}
\Big | \Delta \Big ( \mathbf{e},\mathbf{w}, \frac{1}{\bar{\mathbf{w}}} \Big ) \Big |
& = &
\exp \bigg ( - \sum_{1 \le j < p \le k} \log |e_p - e_j| -
\sum_{j=1}^k \sum_{s=k+1}^{(N+k)/2} \log | w_s - e_j| \Big | {1 \over \bar{w}_s} - e_j \Big | \\
&& - \sum_{k+1 \le a < b \le (N+k)/2} \log | w_b - w_a|
\Big | {1 \over \bar{w}_b} - {1 \over \bar{w}_a} \Big | \bigg ).
\end{eqnarray*}
This is the Boltzmann factor of a two-component log-potential Coulomb gas, consisting of $k$ unit charges at $\{e_j\}_{j=1,\dots,k}$ confined to the unit circle, $(N-k)/2$ unit charges at $\{w_a\}_{a=k+1,\dots,(N+k)/2}$ confined to the unit disk, and a further $(N-k)/2$ image charges to those in the unit disk, which are at positions $\{1/\bar{w}_a\}_{a=k+1,\dots,(N+k)/2}$ outside the unit disk. The other factors in (\ref{eqn:q(y)}) are, from a Coulomb gas perspective, one-body terms due to the coupling of the charges to an external background charge density. We have seen in Proposition \ref{prop:scaled_limit} that in a certain scaled limit the correlation functions for this two-component Coulomb gas tend to functional forms known from the study of the real Ginibre ensemble. These should exhibit features characteristic of a two-component Coulomb gas. Here we will exhibit two such features of the scaled correlations.

In regard to the first of these, suppose that in the limiting system we fix real eigenvalues at $\{x_j\}_{j=1,\dots,k_1}$ and complex eigenvalues at $\{z_j\}_{j=1,\dots,k_2}$ (the latter also requires complex eigenvalues at $\{\bar{z}_j\}_{j=1,\dots,k_2}$). Regarding this action as perturbations, the Coulomb gas perspective tells us that to maintain equilibrium the system will respond by surrounding the fixed eigenvalues with a screening cloud equal and opposite in total charge to that of the perturbation. In terms of the correlations this gives rise to the sum rule (see e.g.~\cite[eq.~(14.20)]{forrester?})
\begin{eqnarray}\label{56c}
&&
\int_{-\infty}^\infty \rho_{(k_1+1,k_2)}^T(\{x_j\}_{j=1,\dots,k_1} \cup \{y\};\{z_j\}_{j=1,\dots,k_2} ) \, dy 
\nonumber \\
&&
\qquad \qquad + 2 \int_{\mathbb R_+^2} \rho_{(k_1,k_2+1)}^T(
\{x_j\}_{j=1,\dots,k_1}; \{z_j\}_{j=1,\dots,k_1} \cup \{z\}) \, d^2 z 
\nonumber \\
&& \qquad =
- (k_1 + 2 k_2) \rho_{(k_1,k_2)}^T(\{x_j\}_{j=1,\dots,k_1}; \{z_j\}_{j=1,\dots,k_2}),
\end{eqnarray} 
where $z=X+iY$, $d^2z = dX dY$, and $\rho_{(k_1,k_2)}^T(\{x_j\}_{j=1,\dots,k_1}; \{z_j\}_{j=1,\dots,k_2})$ is the truncated correlation function
(see \cite[Eq.~(5.3)]{forrester?}). The factor of 2 with $k_2$ in (\ref{56c}) is due to the complex eigenvalues always occurring in complex conjugate pairs.

The explicit form of the $\rho_{(k_1,k_2)}^T$ can be read off from (\ref{eqn:BSlargeN}). For this it is convenient to introduce the quaternion determinant, qdet, according to
$$
{\rm qdet} \, A Z_{2N} = {\rm Pf} \, A, \qquad Z_{2N} :=
\1_N \otimes \begin{bmatrix} 0 & -1 \\ 1 & 0 \end{bmatrix},
$$
for $A$ a $2N \times 2N$ antisymmetric matrix. The correlation function (\ref{eqn:BSlargeN}) then reads
$$
\rho_{(k_1,k_2)}(x,z) =
{\rm qdet}  \begin{bmatrix}[ \tilde{K}_{r,r}(x_j,x_l)]_{j,l=1,\dots,k_1} &
 [\tilde{K}_{r,c}(x_j,z_l)]_{j=1,\dots,k_1 \atop l=1,\dots,k_2} \\
  [\tilde{K}_{c,r}(z_l,x_j)]_{ l=1,\dots,k_2 \atop j=1,\dots,k_1} &
[\tilde{K}_{c,c}(z_j,z_l)]_{j,l=1,\dots,k_2}   \end{bmatrix}
$$
where, with $K_{c,r}(z,x) := - (K_{r,c}(x,z))^T$, each $2 \times 2$ block $\tilde{K}_*(u,v)$ is related to the corresponding block in (\ref{eqn:BSlargeN}) by 
$$
\tilde{K}_*(u,v) = K_*(u,v)
  \begin{bmatrix} 0 & -1 \\ 1 & 0 \end{bmatrix}.
  $$
The advantage of such a  determinant form is that it then follows (see e.g.~\cite[Eq.~(7.184)]{forrester?})  that the corresponding truncated correlations are given by a sum over maximum length cycles in the determinant,
\begin{equation}\label{58}
\rho_{(k_1,k_2)}^T(x,z) = (-1)^{k_1 + k_2 - 1}
\sum_{{\rm cycles} \atop {\rm length \: k_1 + k_2}}
\Big ( \tilde{K}(y_{i_1},y_{i_2})    \tilde{K}(y_{i_2},y_{i_3}) \cdots
 \tilde{K}(y_{i_{k_1 + k_2}},y_{i_1}) \Big )^{(0)}.
 \end{equation}
Here $\{y_i\}_{i=1,\dots,k_1 + k_2} = \{x_i\}_{i=1,\dots,k_1} \cup \{z_j \}_{j=1,\dots,k_2}$, and the operation $( \cdot )^{(0)}$ refers to ${1 \over 2} {\rm Tr}$.
 
The sum rule (\ref{56c}) is a corollary of integration formulas involving the product of two matrix kernels $\tilde{K}_*$.
 
 \begin{proposition}\label{4.5}
 We have
 \begin{eqnarray*}
&&  \int_{-\infty}^\infty \tilde{K}_{r,r}(x,u) \tilde{K}_{r,r}(u,y) \, du +
 2 \int_{\mathbb R_+^2}  \tilde{K}_{r,c}(x,z) \tilde{K}_{c,r}(z,y) \, d^2 z
 \nonumber \\
 && \quad = \tilde{K}_{r,r}(x,y) 
 \begin{bmatrix}0 & 0 \\ 0 & 1 \end{bmatrix} +
 \begin{bmatrix}1 & 0 \\ 0 & 0 \end{bmatrix}\tilde{K}_{r,r}(x,y),
 \end{eqnarray*}
 \begin{eqnarray*}
 && \int_{-\infty}^\infty \tilde{K}_{c,r}(z,y) \tilde{K}_{r,c}(y,v) \, dy +
 2 \int_{\mathbb R_+^2}  \tilde{K}_{c,c}(z,w) \tilde{K}_{c,c}(w,v) \, d^2 w 
 = 2 \tilde{K}_{c,c}(z,v), \\
  && \int_{-\infty}^\infty \tilde{K}_{c,r}(z,y) \tilde{K}_{r,r}(y,x) \, dy +
 2 \int_{\mathbb R_+^2}  \tilde{K}_{c,c}(z,w) \tilde{K}_{c,r}(w,x) \, d^2 w
 =  \tilde{K}_{c,r}(z,x), \\
 && \int_{-\infty}^\infty \tilde{K}_{r,r}(x,y) \tilde{K}_{r,c}(y,z) \, dy +
 2 \int_{\mathbb R_+^2}  \tilde{K}_{r,c}(x,w) \tilde{K}_{c,c}(w,z) \, d^2 w
 =  \tilde{K}_{r,c}(x,z).
  \end{eqnarray*}
  \end{proposition}

\textit{Proof:} Each of the above equations requires evaluating the integrals for the four entries of the matrix products. We will illustrate the required working by giving the details in the case of the $(11)$-component of the first of the equations. With the notation $( X )_{jk}$ denoting the entry $(jk)$ of the matrix $X$, we have
 \begin{eqnarray*}
&& \Big ( \int_{-\infty}^\infty \tilde{K}_{r,r}(x,u) \tilde{K}_{r,r}(u,y) \, du \Big )_{11}
 =  {1 \over 2 \pi} \int_{-\infty}^\infty e^{-{1 \over 2}(x-u)^2 -{1 \over 2}(u-y)^2 } \,du\\
&& \qquad \qquad  + {1 \over 2 \sqrt{2 \pi}}
\int_{-\infty}^\infty (x-u) {\rm sgn}(u-y) e^{-{1 \over 2}(x-u)^2 } {\rm erfc}\Big ( {|u-y| \over
\sqrt{2}} \Big ) \, du.
\end{eqnarray*}
Completing the square shows
$$
{1 \over 2 \pi} \int_{-\infty}^\infty e^{-{1 \over 2}(x-u)^2 -{1 \over 2}(u-y)^2 } \,du
= {1 \over 2 \sqrt{\pi}}  e^{ - {1 \over 4} (x-y)^2},
$$
while writing $(x-u) e^{-{1 \over 2}(x-u)^2} = {\partial \over \partial u}  e^{-{1 \over 2}(x-u)^2} $, integrating by parts and making further use of the above integral evaluation shows
 \begin{eqnarray*}
 &&
 {1 \over 2 \sqrt{2 \pi}}
\int_{-\infty}^\infty (x-u) {\rm sgn}(u-y) e^{-{1 \over 2}(x-u)^2 } {\rm erfc} \,\Big ( {|u-y| \over
\sqrt{2}} \Big ) \, du \\
&& \quad = - {1 \over \sqrt{2 \pi}} e^{-{1 \over 2}(x-y)^2} + {1 \over 2 \sqrt{\pi}}  e^{ - {1 \over 4} (x-y)^2}.
\end{eqnarray*}
Hence
\begin{equation}\label{v1}
\Big ( \int_{-\infty}^\infty \tilde{K}_{r,r}(x,u) \tilde{K}_{r,r}(u,y) \, du \Big )_{11} =
- {1 \over \sqrt{2 \pi}} e^{-{1 \over 2}(x-y)^2}  + {1 \over \sqrt{\pi}}
e^{ - {1 \over 4} (x-y)^2}.
\end{equation}

For the integral over $\mathbb R_+^2$ in the first of the equations, we have
 \begin{eqnarray*}
&& \Big ( \int_{\mathbb R_+^2}  \tilde{K}_{r,c}(x,z) \tilde{K}_{c,r}(z,y) \, d^2 z \Big )_{11}
\\
&& \qquad = {\rm Re} \, {i \over \pi} \int_{\mathbb R_+^2}  {\rm erfc} \,(\sqrt{2} Y)
(\bar{z} - x) e^{-{1 \over 2}(x-\bar{z})^2 -{1 \over 2}(y-z)^2 } \, dX dY.
\end{eqnarray*}
Recalling that $z=X+iY$, completing the square in $X$ and $Y$, translating the integral in $X$, and noting $\int_{-\infty}^\infty X e^{-X^2} \, dX = 0$ reduces this to
$$
{\rm Re} \, {1 \over \pi} \int_{\mathbb R_+^2}  {\rm erfc} \,(\sqrt{2} Y)
\Big (Y - i {x - y \over 2} \Big ) e^{-X^2 + (Y - i {x - y \over 2})^2} \, dX dY.
$$
Now writing $\Big (Y - i {x - y \over 2} \Big ) e^{ (Y - i {x - y \over 2})^2} =
{1 \over 2} {\partial \over \partial Y} e^{(Y - i {x - y \over 2})^2} $ and integrating by parts allows this integral to be evaluated, and we obtain
\begin{equation}\label{v2}
\Big ( \int_{\mathbb R_+^2}  \tilde{K}_{r,c}(x,z) \tilde{K}_{c,r}(z,y) \, d^2 z \Big )_{11}
= - {1 \over 2 \sqrt{\pi}} e^{-(x-y)^2/4} + {1 \over \sqrt{2 \pi}} e^{-(x-y)^2/2}.
\end{equation}
Forming (\ref{v1}) plus twice (\ref{v2}) gives ${1 \over \sqrt{2 \pi}} e^{-(x-y)^2/2}$, which is the (11)-component of $\tilde{K}_{r,r}(x,y)$, in keeping with first of the equations of the proposition.\hfill $\square$

\medskip
To prove the sum rule (\ref{56c}) we substitute (\ref{58}) for the truncated correlations on the LHS of (\ref{56c}). We see that the required integrations can be computed using Proposition \ref{4.5}. When involving the first of the integration formulas
therein, we can either use the cyclic property of the trace, or sum together pairs of terms, to effectively replace the RHS by $\tilde{K}_{r,r}(x,y)$. In all cases this allows the expression resulting from the integration to be identified with a term in the expression for $- \rho_{(k_1,k_2)}^T$ on the RHS of (\ref{56c}) as implied by (\ref{58}). Moreover, each term is repeated $(k_1 + 2k_2)$ times, thus verifying (\ref{56c}).

In addition to the sum rule (\ref{56c}) there is a second sum rule satisfied by the correlation functions, as suggested by the Coulomb gas analogy. For this we consider the complex moments of the screening cloud due to a fixed complex eigenvalue at point $z$. This screening cloud is defined as the function of $w \in \mathbb R_+^2$ and $x \in \mathbb R$ given by
\begin{equation}\label{w6}
2 \rho_{(0,2)}^T(z,w) \chi_{w \in \mathbb R_+^2} +
\rho_{(1,1)}^T(z,x) \chi_{x \in \mathbb R} +
2 \delta^{(2)}(z-w) \rho_{(0,1)}(w).
\end{equation}

We know as a special case of (\ref{56c}) that integrating this over $w$ and $x$ gives zero. Another feature of the screening cloud, seen by inspection of Proposition \ref{prop:BS_limiting_kernels}, is that for large $|w|$ and $|x|$ it decays at a Gaussian rate to 0. In the theory of Coulomb systems (see e.g.~\cite{Ma88}) this rapid decay can be shown to occur only if the complex (multi-pole) moments of the screening cloud all vanish. We know from \cite{Ja82,Fo85} that this statement must be modified when image charges are present so as to relate to the screening cloud of the charge/image system. Consequently, we should re-interpret (\ref{w6}) as the function of $w \in \mathbb R^2\backslash \mathbb R$ and $x \in \mathbb{R}$ specified by
\begin{equation}\label{w7}
 \rho_{(0,2)}^T(z,w) \chi_{w \in \mathbb R_+^2}+
\rho_{(1,1)}^T(z,x) \chi_{x \in \mathbb R} +
 \delta^{(2)}(z-w) \rho_{(0,1)}(w),
\end{equation}
with the property that
\begin{equation}\label{w8}
 \rho_{(0,2)}^T(z,\bar{w}) = \rho_{(0,2)}^T(z,w) \qquad
 \rho_{(0,1)}(w) = \rho_{(0,1)}(\bar{w}).
\end{equation}

By translation invariance of the system in the $x$-direction we can set $z=iy_0$. We then see that with the conditions (\ref{w8}) the odd complex moments of (\ref{w7}) vanish by symmetry, whereas the vanishing of the even complex moments requires
$$
2 \int_{\mathbb R^2_+} w^{2p} \rho_{(0,2)}^T(z,w) \, d^2 w +
\int_{-\infty}^\infty x^{2p}  \rho_{(1,1)}^T(z,x) \, dx = - 2 z^{2p} \rho_{(0,1)}(z) \quad p=0,1,\dots
$$
Recalling the explicit form of the truncated correlations as implied by (\ref{58}) and Proposition \ref{prop:BS_limiting_kernels}, by multiplying both sides by $\alpha^p/p!$ and summing over $p$ we obtain an equivalent form of this sum rule, which we state and prove in the following result.

\begin{proposition}
Let $z=iy_0$ and suppose $|\alpha| < 1$. We have
$$
2 \int_{\mathbb R^2_+}  e^{\alpha w^2} \tilde{K}_{c,c}(z,w) \tilde{K}_{c,c}(w,z) \,
d^2 w +
\int_{-\infty}^\infty e^{\alpha x^2} \tilde{K}_{c,r}(z,x) \tilde{K}_{r,c}(x,z) \, dx =
2 e^{\alpha z^2} \tilde{K}_{c,c}(z,z).
$$
\end{proposition}

\textit{Proof:}  We will illustrate our methods by considering the (11)-component. For the first term on the LHS this component can be written
\begin{eqnarray*}
&&{1 \over 4 \pi} {\rm erfc} \,(\sqrt{2} y_0)
\int_0^\infty dY \int_{-\infty}^\infty dX \, e^{\alpha (X + i Y)^2}
{\rm erfc} (\sqrt{2} Y) \\
&& \qquad \times 
\Big ( {\partial^2 \over \partial X^2} + {\partial^2 \over \partial Y^2} \Big )
\Big ( ( e^{(Y+y_0)^2} - e^{(Y - y_0)^2} ) e^{-X^2} \Big ).
\end{eqnarray*}
Separating the terms involving the partial derivatives, and integrating by parts in each reduces the double integral to
\begin{eqnarray*}
&& - 4 y_0 e^{y_0^2} \int_{-\infty}^\infty e^{\alpha X^2 - X^2} \,dX \\
&& - {8 \sqrt{2} \over \pi} \int_0^\infty dY \int_{-\infty}^\infty dX \,
\Big ( i \alpha(X+iY) - Y \Big ) e^{\alpha(X+iY)^2 - 2Y^2 - X^2}
( e^{(Y+y_0)^2} - e^{(Y-y_0)^2} ).
\end{eqnarray*}
Upon completing the square in $X$, simplifying, then completing the square in $Y$  the second of these integrals can be evaluated, giving that the (11)-component of the first term on the LHS is equal to
$$
{1 \over 4 \pi} {\rm erfc} \,(\sqrt{2} y_0)
\Big ( - 4y_0 e^{y_0^2} \int_{-\infty}^\infty e^{-(1 - \alpha) X^2} \, dX
+ 8 \sqrt{2 \pi} e^{2y_0^2 - \alpha y_0^2} y_0 \Big ).
$$
An analogous strategy in relation to the (11)-component of the second term on the LHS shows that it is equal to
$$
{1 \over \pi} {\rm erfc} \,(\sqrt{2} y_0) y_0 e^{y_0^2} \int_{-\infty}^\infty e^{-(1 - \alpha) X^2} \, dX.
$$

Adding together the above evaluations of the (11)-components of the terms on the LHS gives
$$
e^{-\alpha y_0^2} \Big ( 2 \sqrt{2 \over \pi}  {\rm erfc} \,(\sqrt{2} y_0) y_0 e^{2y_0^2} \Big ),
$$
which we recognize as the (11)-component of $2e^{-\alpha y_0^2} \tilde{K}_{c,c}(z,z)$, as required by the RHS of the sum rule. \hfill $\square$

\appendix
\makeatletter
\def\@seccntformat#1{\csname Pref@#1\endcsname \csname the#1\endcsname\quad}
\def\Pref@section{Appendix~}
\makeatother
\renewcommand{\thetheorem}{\Roman{theorem}}
\section{}
\label{app:PY}

The following lemmata from \cite{muirhead1982} are required in the proof of Proposition \ref{prop:elementjpdf}.

\begin{lemma}[Theorem 2.1.5]
\label{lem:alpha_tensor_beta}
For $X=\alpha Y \beta$, where $\alpha_{P\times P}$ and $\beta_{Q\times Q}$ are arbitrary real matrices and $Y_{P\times Q}$ has $PQ$ independent entries (ie. the wedge product $(dY)$ has $PQ$ factors) then
\begin{eqnarray*}
(dX)&=&\left|\mathrm{det}(\alpha \otimes\beta^T)\right|(dY)\\
&=&\left|\mathrm{det}(\alpha)^Q\mathrm{det}(\beta)^P\right|(dY).
\end{eqnarray*} 
\end{lemma}

\begin{lemma}[Theorem 2.1.14]
\label{lem:tilde_const_covarbs}
For any $N\times M$ $(M\geq N)$ matrix $X$, if $W=XX^T$ then
\begin{eqnarray}
\label{eqn:tilde_const_covarbs} (dX)= c \hspace{3pt}\mathrm{det}W^{(N-M-1)/2}(dW),
\end{eqnarray}
where $c$ is independent of $W$.
\end{lemma}

\begin{lemma} [Theorem 2.1.6]
\label{lem:adma}
For $A$ an $N \times N$ real non-singular matrix and $M$, an $N \times N$ real symmetric matrix, one has
\begin{eqnarray}
\nonumber (A^TdMA)&=&\mathrm{det}(A^TA)^{(N-1)/2+1}(dM).
\end{eqnarray}
\end{lemma}

\begin{lemma}[Ch.3, Equation (22)]
\label{lem:beta=1jacobian}
For $C$ a real symmetric $N \times N$ matrix
\begin{eqnarray}
\nonumber (dC)=\prod_{j<k}^N (x_k-x_j) dx_1\cdot\cdot\cdot dx_N (R^TdR),
\end{eqnarray}
where the $\{x_i\}$ are the ordered eigenvalues of $C$, and $R$ is the real orthogonal matrix of eigenvectors.
\end{lemma}

\subsection{Proof of Proposition \ref{prop:elementjpdf}}

Recall that $Y=A^{-1}B$. By letting $\alpha=A$ and $\beta=\mathbf{1}_{N}$ in Lemma \ref{lem:alpha_tensor_beta} we see that
\begin{eqnarray}
\nonumber \label{dBi}(dB)&=&|\mathrm{det} A|^N(dY),
\end{eqnarray}
so we rewrite (\ref{eqn:YjpdfAB}) as
\begin{eqnarray}
\nonumber \mathcal{P}(Y)(dA)(dB)&=&(2\pi)^{-N^2}e^{-\frac{1}{2}\mathrm{Tr}\left(AA^T(\mathbf{1}_N+YY^T)\right)}|\mathrm{det}AA^T|^{N/2}(dA)(dY).
\end{eqnarray}
(Here and below $(dA)(dB)$, and similar, is to be interpreted as the wedge product of the corresponding differentials.) Setting $C:=AA^T$, Lemma \ref{lem:tilde_const_covarbs} tells us that $(dA)=c (\det C)^{1/2}(dC)$. Integrating over $C$ (noting that $C$ is positive definite, denoted $C>0$) we have
\begin{eqnarray}
\nonumber \mathcal{P}(Y)(dY)&=&(2\pi)^{-N^2} c\int_{C>0}(\mathrm{det}C)^{(N-1)/2}e^{-\frac{1}{2}\mathrm{Tr}\left(C(\mathbf{1}_N+YY^T)\right)}(dC)(dY)\\
\nonumber &=&(2\pi)^{-N^2} c\int_{C>0}(\mathrm{det}C)^{(N-1)/2}e^{-\frac{1}{2}\mathrm{Tr}\left((\mathbf{1}_N+YY^T)^{1/2}C(\mathbf{1}_N+YY^T)^{1/2}\right)}(dC)(dY).
\end{eqnarray}
Carrying out the change of variables $C\rightarrow (\mathbf{1}_N+YY^T)^{1/2}C(\mathbf{1}_N+YY^T)^{1/2}$ we use Lemma \ref{lem:adma} to find
\begin{eqnarray} 
\nonumber \mathcal{P}(Y)(dY)&=&(2\pi)^{-N^2}c \hspace{3pt}\mathrm{det}(\mathbf{1}_N+YY^T)^{-N}\int_{C>0}\mathrm{det}(C)^{(N-1)/2}e^{-\frac{1}{2}\mathrm{Tr}(C)}(dC)(dY).
\end{eqnarray}
Using Lemma \ref{lem:tilde_const_covarbs} we can calculate $c$ according to
\begin{eqnarray}
\nonumber \int e^{-\mathrm{Tr}(AA^T)/2}(dA)&=&c \int_{C>0}e^{-\mathrm{Tr}(C)/2}\frac{(dC)}{\mathrm{det}(C)^{1/2}},
\end{eqnarray}
and so
\begin{eqnarray}
\nonumber c&=&\frac{(2\pi)^{N^2/2}}{\int_{C>0}\mathrm{det}(C)^{-1/2}e^{-\mathrm{Tr}(C)/2}(dC)},
\end{eqnarray}
telling us that
\begin{eqnarray}
\nonumber \mathcal{P}(Y)(dY)&=&(2\pi)^{-N^2/2}\mathrm{det}(\mathbf{1}_N+YY^T)^{-N}\frac{\int_{C>0}\mathrm{det}(C)^{(N-1)/2}e^{-\mathrm{Tr}(C)/2}(dC)}{\int_{C>0}\mathrm{det}(C)^{-1/2}e^{-\mathrm{Tr}(C)/2}(dC)}(dY).
\end{eqnarray}
Since $C=AA^T$ is symmetric, using Lemma \ref{lem:beta=1jacobian} the ratio of integrals can be rewritten as
\begin{eqnarray}
\nonumber &&\frac{\int_{C>0}\mathrm{det}(C)^{(N-1)/2}e^{-\frac{1}{2}\mathrm{Tr}(C)}(dC)}{\int_{C>0}(\mathrm{det}C)^{-1/2}e^{-\mathrm{Tr}(C)/2}(dC)}\\
\nonumber \label{eqn:ratio_selberg_ints}&& =  \frac{\int_{(0,\infty)^N}\prod_{l=1}^Nx_l^{(N-1)/2}e^{-x_l/2}\prod_{j<k}^N|x_k-x_j|dx_1\cdot\cdot\cdot dx_N}{\int_{(0,\infty)^N}\prod_{l=1}^Nx_l^{-1/2}e^{-x_l/2}\prod_{j<k}^N|x_k-x_j|dx_1\cdot\cdot\cdot dx_N},
\end{eqnarray}
which is seen to be a ratio of Selberg-type integrals, which have known evaluations 
in terms of gamma functions (see e.g.~\cite[Ch. 4]{forrester?}) . The result now follows.
\hfill $\Box$

\section{}
\label{app:2}

The purpose of this appendix is to integrate over $\tilde{R}_N$, the strictly upper triangular elements of $R_N$ in (\ref{eqn:RN}), leaving us with just the dependence on the eigenvalues. This will be done column-by-column, starting with the case where $j>k$, that is, the columns corresponding to the complex eigenvalues, and then proceeding onto those columns corresponding to the real eigenvalues.

\subsection{Complex eigenvalue columns}

In the region $j>k$ of $R_N$ we can isolate the last two rows and columns to write
\begin{eqnarray}
\nonumber R_N=\left[\begin{array}{cc}
R_{N-2} & u\\
0^T & z_m
\end{array}\right],
\end{eqnarray}
where $u$ is of size $(N-2)\times 2$ and $0^T$ is of size $2\times (N-2)$. So then
\begin{eqnarray}
\nonumber \1_N +R_NR_N^T&=&\left[\begin{array}{cc}
\1_{N-2}+R_{N-2}R_{N-2}^T+uu^T -uz_m^T(\1_2+z_mz_m^T)^{-1}z_mu^T & 0\\ 
z_mu^T & \1_2 +z_mz_m^T
\end{array}\right]
\end{eqnarray}
and
\begin{eqnarray}
\label{eqn:det1+rr}
\nonumber \det(\1_N+R_NR_N^T)&=&\det(\1_2+z_mz_m^T)\det(\1_{N-2}+R_{N-2}R_{N-2}^T)\\
\nonumber \label{eqn:det1+RnRn}&&\times\det(\1_2+(\1_2+z_mz_m^T)^{-1}u^T(\1_{N-2}+R_{N-2}R_{N-2}^T)^{-1}u),
\end{eqnarray}
where we have used the general identity (for appropriate sized $\mathbf{1}$)
\begin{eqnarray}
\label{eqn:1+ab}
\det(\mathbf{1}+AB)&=&\det(\mathbf{1}+BA).
\end{eqnarray}

We are now in a position to integrate over the elements of the matrix $u$
{\small
\begin{eqnarray}
\nonumber &&\int\frac{(du)}{\det(\1_N+R_NR_N^T)^N}=\frac{1}{\det(\1_2+z_mz_m^T)^N\det(\1_{N-2}+R_{N-2}R_{N-2}^T)^N}\\
\nonumber &&\times \int\frac{(du)}{\det(\1_{2}+(\1_2+z_mz_m^T)^{-1/2}u^T(\1_{N-2}+R_{N-2}R_{N-2}^T)^{-1}u(\1_2+z_mz_m^T)^{-1/2})^N},
\end{eqnarray}
}

\noindent where the integral for each independent real component of $u$ is over the real line. Changing variables $ v=(\1_{N-2}+R_{N-2}R_{N-2}^T)^{-1/2}u(\1_2+z_mz_m^T)^{-1/2}$ we use Lemma \ref{lem:alpha_tensor_beta} to find
\begin{eqnarray}
\nonumber \int\frac{(du)}{\det(\1_N+R_NR_N^T)^N}&=&\frac{1}{\det(\1_2+ z_mz_m^T)^{N/2+1} \det(\1_{N-2} +R_{N-2}R_{N-2}^T)^{N-1}}\\
\nonumber &&\times\int\frac{(dv)}{\det(\1_{2}+v^Tv)^N}.
\end{eqnarray}
Iterating over all columns corresponding to complex eigenvalues we have
\begin{align}
\nonumber &\int\frac{(du_{N-2})\cdot\cdot\cdot(du_{k+1})}{\det(\1_N+R_NR_N^T)^N}=\frac{1}{\det(\1_k+R_kR_k^T)^{(N+k)/2}}\\
\label{eqn:with_subs} &\quad \times\prod_{s=k+1}^{(N+k)/2}\frac{1}{\det(\1_2+z_sz_s^T)^{N/2+1}}\prod_{s=0}^{(N-k)/2-1}\int\frac{(dv_{N-2-2s})}{\det(\1_2+v_{N-2-2s}^Tv_{N-2-2s})^{N-s}},
\end{align}where the subscripts $*$ on the matrices $v_*,du_*,dv_*$ denote their number of rows.

To evaluate each of the $(N-k)/2$ integrals we use a similar method to that used in Proposition \ref{prop:elementjpdf}. Firstly, for each $v_{N-2-2s}$, we let $v_{N-2-2s}^Tv_{N-2-2s}=C$ and apply Lemma \ref{lem:tilde_const_covarbs} to get 
\begin{equation}\label{11'}
(dv)=c (\det C)^{(N-2s-5)/2}(dC)
\end{equation}
 and 
\begin{eqnarray}
\nonumber c \int (\det C)^{(N-2s-5)/2}e^{-\Tr (C)}(dC)=\int e^{-\Tr(v^Tv)}(dv)=\pi^{N-2s-2}.
\end{eqnarray}
And so, with $\kappa:=(N-2s-5)/2$,
\begin{eqnarray}
&&\nonumber \int\frac{(dv_{N-2-2s})}{\det(\1_2+v_{N-2-2s}^Tv_{N-2-2s})^{N-s}}=\pi^{N-2s-2}\frac{\int(\det C)^{\kappa}\det(\1_2+C)^{s-N}(dC)}{\int(\det C)^{\kappa}e^{-\Tr C}(dC)}\\
&&\nonumber  \quad = \pi^{N-2s-2}\int_0^{\infty}\int_0^{\infty}\frac{x_1^{\kappa}}{(1+x_1)^{N-s}}\frac{x_2^{\kappa}}{(1+x_2)^{N-s}}|x_1-x_2|dx_1dx_2\\
\nonumber &&\qquad \times\left( \int_0^{\infty}\int_0^{\infty}x_1^{\kappa}x_2^{\kappa}e^{-x_1}e^{-x_2} |x_1-x_2|dx_1dx_2  \right)^{-1},\\
&&\quad \nonumber =\pi^{N-2s-2}\int_0^1\int_0^1y_1^{\kappa}y_2^{\kappa}(1-y_1)^{(N-1)/2}(1-y_2)^{(N-1)/2}|y_1-y_2|dy_1dy_2\\
\label{eqn:compx_selberg} &&\qquad \times\left( \int_0^{\infty}\int_0^{\infty}x_1^{\kappa}x_2^{\kappa}e^{-x_1}e^{-x_2} |x_1-x_2|dx_1dx_2  \right)^{-1},
\end{eqnarray}
where use was made of Lemma \ref{lem:beta=1jacobian} for the second equality, and the change of variables $y=x/(1+x)$ for the third.

A ratio of Selberg integrals has again appeared in (\ref{eqn:compx_selberg}), and using results from \cite{forrester?} we find
\begin{eqnarray}\label{12'}
\int\det(\1_2+v_{N-2-2s}^Tv_{N-2-2s})^{-(N-s)}(dv)=\pi^{N-2s-2}\frac{\Gamma((N+1)/2)}{\Gamma(N-s-1/2)}\frac{\Gamma(N/2+1)}{\Gamma(N-s)}.
\end{eqnarray}
The case $N$ odd, $k=1$, corresponding to $s=(N-1)/2-1$ is special since then $v_{N-2-2s}$
consists of 1 row and 2 columns, and thus is the only case in which the number of rows is less than the
number of columns. We must then write
$$
\det ({\bf 1}_2 + v^T_{N-2-2s} v_{N-2-2s})^{-p} = (1 + v_{N-2-2s} v_{N-2-2s}^T)^{-p},
$$
using (\ref{eqn:1+ab}). However, it turns out that the change this implies to 
(\ref{eqn:compx_selberg}) does not effect the evaluation (\ref{12'}), even though (\ref{11'}) is no longer valid.
So in all cases, after having integrated over the $\tilde{R}_{i,j}$ for $i>j$ in the columns corresponding to complex eigenvalues we are left with
\begin{eqnarray}
\nonumber &&\int\frac{(du_{N-2})\cdot\cdot\cdot(du_{k+1})}{\det(\1_N+R_NR_N^T)^N}=\prod_{s=k+1}^{(N+k)/2}\frac{1}{\det(\1_2+z_sz_s^T)^{N/2+1}}\\
\nonumber &&\qquad \times\prod_{s=0}^{(N-k)/2-1}\pi^{N-2s-2}\frac{\Gamma((N+1)/2)}{\Gamma(N-s-1/2)}\frac{\Gamma(N/2+1)}{\Gamma(N-s)} \hspace{3pt}\frac{1}{\det(\mathbf{1}_k+R_kR_k^T)^{(N+k)/2}}.
\end{eqnarray}

It remains to compute the integrals over the columns corresponding to the real eigenvalues.

\subsection{Real eigenvalue columns}

We see that we are left with a function of $R_k$, which is the upper-left sub-block of $R_N$ and we isolate the last row and column
\begin{eqnarray}
\nonumber R_k=\left[\begin{array}{cc}
R_{k-1} & u_{k-1}\\
0^T & \lambda_k
\end{array}\right],
\end{eqnarray}
where now $u_{k-1}$ is of size $(k-1)\times 1$ and $0^T$ is of size $1\times (k-1)$. Following the same procedure as for the complex eigenvalue columns, we find
\begin{eqnarray}
\nonumber \det(\1_k+R_kR_k^T)&=&(1+\lambda_k^2)\det(\1_{k-1}+R_{k-1}R_{k-1}^T)\\
\nonumber &&\times(1+(1+\lambda_k^2)^{-1}u_{k-1}^T(\1_{k-1}+R_{k-1}R_{k-1}^T)^{-1}u_{k-1}).
\end{eqnarray}
Setting $v_{k-1}=(\1_{j-1}+R_{j-1}R_{j-1}^T)^{-1/2} u_{k-1}(1+\lambda_j^2)^{-1/2}$ and again making use of Lemma \ref{lem:alpha_tensor_beta} we have
\begin{eqnarray}
\nonumber \int\frac{(du_{k-1} )}{\det(\1_k+R_kR_k^T)^{(N+k)/2}}&= &\frac{1}{(1+\lambda_k^2)^{(N+1)/2} \det(\1_{k-1}+R_{k-1}R_{k-1}^T)^{(N+k-1)/2}}\\
\nonumber &&\times\int\frac{(dv_{k-1})}{(1+v_{k-1}^T v_{k-1})^{(N+k)/2}}.
\end{eqnarray}
Iterating over the remaining columns of $R_k$ gives
\begin{eqnarray}
\nonumber \int\frac{(du_{k-1})\cdot\cdot\cdot(du_1)}{\det(\1_k +R_kR_k^T)^{(N+k)/2}}=\prod_{s=1}^k\frac{1}{(1+\lambda_s^2)^{(N+1)/2}}\prod_{s=1}^{k-1}\int\frac{(dv_{k-s})}{(1+v_{k-s}^Tv_{k-s})^{(N+k)/2-(s-1)/2}}
\end{eqnarray}
(cf.~(\ref{eqn:with_subs})). To evaluate the integrals, we use the same method as for the integrals in (\ref{eqn:with_subs}) (involving Lemma \ref{lem:tilde_const_covarbs} and now one-dimensional case of the Selberg integral, which is the beta integral). This gives
\begin{eqnarray}
\nonumber \int\frac{(dv_{k-s})}{(1+v_{k-s}^Tv_{k-s})^{(N+k)/2-(s-1)/2}}=\pi^{(k-s)/2}\frac{\Gamma((N+1)/2)}{\Gamma((N+k-s+1)/2)},
\end{eqnarray}
and so
\begin{eqnarray}
\nonumber \int\frac{(du_{k-1})\cdot\cdot\cdot(du_1)}{\det(\1_k+R_kR_k^T)^{(N+k)/2}}=\prod_{s=1}^k\frac{1}{(1+\lambda_s^2)^{(N+1)/2}}\prod_{s=1}^{k-1}\pi^{(k-s)/2}\frac{\Gamma((N+1)/2)}{\Gamma((N+k-s+1)/2)}.
\end{eqnarray}

\subsection*{Acknowledgements}
The work of PJF was supported by the Australian Research Council, and AM was supported
by an Australian Postgraduate Award. We thank Dan Mathews for bringing up the topic of
random tensors during a discussion at the 1st PRIMA meeting (Sydney, July 2009). Discussions with J. Fischmann in relation to (\ref{eqn:charpolydensity}) are acknowledged.

\end{document}